\documentclass[]{aa}
\usepackage{color}
\usepackage{amsmath}
\usepackage{mathrsfs}
\usepackage{graphicx}
\usepackage{natbib}
\bibpunct{(}{)}{;}{a}{}{,}
\newcommand{\XSPECB}{\texttt{Xspec}}
\def\compmag{{\sc compmag}}
\def\phabs{{\sc phabs}}

\def\pcfabs{{\sc pcfabs}}

\def\highecut{{\sc highecut}}
\def\diskbb{{\sc diskbb}}

\def\highecut{{\sc highecut}}
\def\sax{\textsl{BeppoSAX}\xspace}

\def\suzaku{\textsl{Suzaku}\xspace}
\def\nustar{\textsl{NuStar}\xspace}

\def\zmax{z_{\rm max}}

\def\sigmacyc{\sigma_{\rm cyc}}

\def\sigmag{\sigma_{\rm G}}

\def\scw{Schwarzschild}
\def\zmax{z_{\rm max}}
\def\ktbb{kT_{\rm bb}}
\def\kte{kT_{\mathrm e}}
\def\ecyc{E_{\rm cyc}}

\def\ne{n_{\mathrm e}}
\def\me{m_{\mathrm e}}
\def\Nh{N_{\mathrm H}}
\def\ncomp{N_{\rm comp}}
\def\mdot{\dot{m}}
\def\fu{4U~0115$+$63}
\def\her{Her~X-1\xspace}
\def\cen{Cen~X-3\xspace}

\def\b12{B_{\rm 12}}
\def\bs{bremsstrahlung}
\def\fgeom{f_{\rm geom}}
\def\chiq{$\chi^2$}
\def\uf{u_{\rm f}}
\def\ul{u_{\rm l}}
\def\ef{E_{\rm f}}
\def\el{E_{\rm l}}

\def\mz{M^{\rm z}}
\def\aff{\alpha_{\rm ff}}

\def\xspec{\texttt{Xspec}\xspace}
\newcommand{\lecs}{\textsl{LECS}\xspace}
\newcommand{\mecs}{\textsl{MECS}\xspace}
\newcommand{\pds}{\textsl{PDS}\xspace}
\newcommand{\hp}{\textsl{HPGSPC}\xspace}

\begin{document}

\title{A new model for the X-ray continuum of the magnetized accreting pulsars}

\author{Ruben Farinelli\inst{1}
	\and
	Carlo Ferrigno\inst{2}
        \and
        Enrico Bozzo\inst{2}
 			\and
        Peter A. Becker\inst{3}       
}

\authorrunning{R. Farinelli et al.}
\titlerunning{X-ray spectra of accreting pulsars}
   \offprints{R. Farinelli}
\institute{$^{1}$INAF-Osservatorio Astronomico di Padova, Vicolo dell'Osservatorio 5, I-35122 Padova, Italy\\
	\email{ruben.farinelli@oapd.inaf.it}\\
	$^{2}$ISDC, chemin d'\'Ecogia 16, 1290 Versoix, Switzerland\\
	$^{3}$Department of Physics and Astronomy, George Mason University, Fairfax, VA 22030\\
          }

\date{Received --; accepted --}

\abstract
	{Accreting highly magnetized pulsars in binary systems are among the brightest X-ray emitters in our Galaxy. Although a number of 
	high statistical quality broad-band (0.1-100\,keV) X-ray observations are available, the spectral energy distribution of 
	these sources is usually investigated by adopting pure phenomenological models, rather than models linked to the physics 
	of accretion.} 
	{In this paper, a detailed spectral study of the X-ray emission 
	recorded from the high-mass X-ray binary pulsars  \cen, \fu,\ and \her\ is carried out by using  
	\sax\ and joined \suzaku+\nustar data, together with an advanced version of the \compmag\ model. 
         The latter provides a physical description of the high energy emission from accreting pulsars, including 
         the thermal and bulk Comptonization of cyclotron and \bs\ seed photons 
         along the neutron star accretion column.} 
         {The \compmag\ model is based on an iterative method for solving second-order partial differential equations, 
         whose convergence algorithm has been improved and consolidated during the preparation of this paper.} 
        {Our analysis shows that the broad-band X-ray continuum of all considered sources can be self-consistently described by 
        the \compmag\ model. The cyclotron absorption features, not included in the model, can be accounted for by using 
        Gaussian components. From the fits of the \compmag\ model to the data we inferred the physical properties 
        of the accretion columns in all sources, finding values reasonably close to those theoretically expected according to our current 
        understanding of accretion in highly magnetized neutron stars.}  
	{The updated version of the \compmag\ model has been tailored to the physical processes that are known to occur in the columns 
	of highly magnetized accreting neutron stars and it can thus provide a better understanding of the high energy radiation 
	from these sources. The availability of broad-band high statistical quality X-ray data, as those provided by \sax\ in the past 
	and \nustar\ plus other facilities in the present, is crucial to fully exploit the potentialities of the 
	model. The close advent of the Astro-H mission, endowed with an unprecedented combination of high sensitivity and X-ray broad-band 
	coverage, provides good perspectives to improve our understanding of accretion onto highly magnetized neutron star 
	through physical models like the one adopted here.} 


\maketitle

\section{Introduction}
\label{sec:intro} 

X-ray binary pulsars (XRBPs) were discovered more than forty years ago
with the pioneering observations of the bright sources Her~X-1
\citep{Giacconi1971} and Cen~X-3 \citep{tananbaum1972}. The origin 
of the pulsed emission was soon understood to be related to the accretion 
of the material lost by the companion onto a rotating neutron star (NS).  
If the latter is endowed with a sufficiently strong magnetic field 
($B\sim10^{11-12}$\,G), the inflowing material from the donor star 
is halted at the magnetospheric boundary of the compact object and 
funneled toward its magnetic poles, forming one or more accretion columns
\citep{Pringle72,Davids73}. As the NS magnetic field is known to decay 
over time, accretion columns are expected to be more extended in 
relatively young binary systems, as the high mass X-ray binaries 
\citep[HMXB; see, e.g.,][for a recent review]{Walter15}. Some binary systems  
with intermediate properties between HMXBs and low mass X-ray binaries (LMXBs) 
also show evidence for extended accretion columns \citep[IMXBs; see, e.g., 
the case of Her\,X-1;][hereafter BW07]{becker2007}. In all these cases, 
the gravitational potential energy of the accreting material 
is first converted into kinetic energy within the accretion column, and 
then released in the form of X-rays as the plasma decelerates and settles 
onto the stellar surface. X-ray pulsations are generated owing to the disalignment 
between the NS magnetic and rotational axes. 
A peculiar feature that is observed in the X-ray spectra of many HMXBs and IMXBs is the 
cyclotron resonant scattering absorption line (CRSFs), which provide a 
direct measurement of the NS magnetic-field strength. The 
latter can be estimated by using the relation  
$E_{\rm cyc} = 11.6\,B_{12}\times (1+z)^{-1}$\,keV, where 
$E_{\rm cyc}$ is the centroid energy of the fundamental CRSF, 
$B_{12}$ is the NS magnetic field strength in units of $10^{12}$\,G, and $z$ is the
gravitational redshift in the line-forming region \citep{wasserman1983}.
In some cases, higher order harmonics of the fundamental CRSF are also observed 
at higher energies \citep[see, e.g.][and references therein]{Walter15}.  

The continuum broad-band X-ray emission of accreting X-ray pulsars is usually 
described by using phenomenological models, including an absorbed power-law extending up to 
$\sim$100~keV with a roll-over at $\sim$30-50~keV or a broken 
power-law \citep{orlandini06}. In some cases it has been shown, however, that such simplified 
models are no able to satisfactorily describe high statistical quality observations \citep[][hereafter F09]{ferrigno09}. 
The development of refined spectral models, linking the emission from X-ray pulsars to 
the physics of accretion, has been limited so far by the complexity of the radiative and dynamical 
equations that describe the behaviour of the accreting material under extreme gravity and 
magnetic field conditions. BW07 made a major step forward in this respect by elaborating an analytical 
approximation of the theoretical model for accretion onto highly magnetized NSs  
proposed previously by \citet{Davids73}. 

Assuming the case of constant temperature and magnetic field in the 
emitting region, together with a simplified profile for the velocity of the radiation dominated flow in the accretion column, 
BW07 found an analytical solution to solve the Compton reprocessing of seed photons in this region 
(the latter being emitted either at the base of the accretion column or along the accretion stream). 
Through the definition of a proper Green function, these authors managed
to approximate the complex magnetic \bs\ emission
of the electrons in the accretion flow \citep[hereafter]{riffert1999} as a combination of a line emission at 
the cyclotron energy and an ordinary \bs\ emission at lower energies. The angular-dependent 
cross section of the magnetic Compton scattering was then computed by using weighted averages. 
The BW07 model was first revised by \citet[][hereafter F12]{farinelli2012a}, relaxing the previous assumption on the velocity 
profile of the accreting material in the accretion column. These authors considered the case of 
a free-falling plasma settling onto the NS surface, and carried out the integration 
of the Green function along the accretion column. They limited their analysis to the case of low luminosity 
X-ray pulsars \citep[e.g., the supergiant fast X-ray transients in quiescence;][]{Walter15}, such that the bulk of 
the seed photons to be Comptonized can be assumed to originate only from the blackbody emission at the base of the column.

In this work, we further extended the revised approach of F12 
by including in the Comptonization process the seed photons produced by the \bs\ 
and cyclotron emissions along the accretion column. We also allow the value of the dipolar  
magnetic field to vary along the cylindrical column, thus exploring the more realistic situation 
in which the energy and the intensity of the cyclotron emission are directly linked 
to the height of the accretion column and the local density of the plasma. This improvement is motivated by 
the results of F09, who applied the BW07 model to the emission of
the Be/X-ray binary \fu\ and found that the cyclotron emission is characterized by a
magnetic field lower than the one estimated from the centroid energy of the fundamental CSRF. 
They suggested that this discrepancy might have been caused by a difference in the height of the regions
producing the continuum cyclotron emission and the scattering feature along the accretion column. 

In order to test the presently improved model, we used a number of observations collected with 
\sax\ and the joined \suzaku\ + \nustar\ facilities of the three sources \her, \cen, and \fu,\ 
which are well known to possess prominent cyclotron lines in their X-ray spectra. 
The X-ray spectrum of \fu\ is also known to be characterized by the presence of a very strong and 
broad emission-like feature at 10\,keV, i.e. right below the energy of the fundamental cyclotron absorption line (F09). 
This ``10-keV feature'' \citep{coburn2002} has been observed in a number of X-ray pulsars, and it is commonly modelled 
in the literature by using a broad Gaussian line \citep[F09;][]{suchy2008, vasco2013} which origin is not well understood.  
In our model, we show that the 10-keV feature can be reasonably well explained as 
the emission from the collisionally excited first Landau level (cyclotron emission) broadened by the Comptonization 
along the accretion column. 

We present an overview of the improved \compmag\ model in Sect.~\ref{sec:new_compmag}, 
and summarize in Sect.~\ref{sec:sources} the properties of the X-ray pulsars 
\her, \cen, and \fu\ for which we carry out a detailed spectral analysis to test the model. 
In Sect.~\ref{sec:observation}, we describe our data-set and provide some details 
on the data reduction procedures. All results are summarized in Sect.~\ref{sec:results} and 
discussed in Sect.~\ref{sec:discussion}. The theoretical considerations for the seed
photons emission terms and the accretion geometry are discussed in Sect.~\ref{sec:source_terms} and
and Sect.~\ref{sec:geom}, respectively. We provide our conclusions in Sect.~\ref{sec:conclusions}.

\section{The updated \compmag\ model}
\label{sec:new_compmag}

The first version of the \compmag\ model was presented by F12, and the 
main difference with respect to the spectral model developed by BW07 was 
related to the presence of the second-order bulk-Comptonization term in the energy diffusion space 
operator, together with the possibility to consider a more general velocity profile for the inflowing matter 
along the pulsar accretion column. The latter was approximated as cylindrical and permeated by a constant magnetic field. 
The seed photons for the Comptonization were assumed to be uniquely from the blackbody  radiation at the bottom 
of the column and diffused through its height. 
The algorithm implemented by F12 to solve the Fokker-Planck approximation of 
the radiative transfer equation in the accretion column is based on an general finite-difference iterative procedure for the 
solution of partial differential equations with two variables. 

The first version of the \compmag\ model was applied to the case of the Supergiant Fast X-ray Transients, 
as these sources host pulsars which display usually a relatively low X-ray luminosity 
\citep[$\simeq 10^{33-34}$~erg~s$^{-1}$;][]{farinelli2012b}. As shown by BW07, in brighter accreting X-ray 
pulsars the contribution of the Comptonized blackbody is virtually negligible compared to that of the Comptonized \bs\ 
and the cyclotron emission. The latter thus need to be considered if bright X-ray pulsars are studied.  
This extension is presented in this work and represents the key difference between the previous and the updated 
version of the \compmag\ model. 

 For electrons embedded in a strong magnetic field, the motion in the direction
perpendicular to field lines has discrete energy states (the Landau levels),
while along the field lines the velocity is distributed according
to a function $f(v)$. The electron distribution function $f(v)$ can be determined 
from the solution of the integro-differential Boltzmann equation with a particle diffusion term
and a Coulomb scattering  kernel \citep[][hereafter R99]{riffert1999}.
The proton-electron Coulomb interaction without photon
emission tends to bring the electrons to a thermodynamical equilibrium
situation, for which $f(v)$ is a Maxwellian at the proton temperature $T$.
If the electron velocity is such that $v > \sqrt{2 \ecyc/\me}$, where $\ecyc$ is the cyclotron energy, 
the Coulomb scattering cross-section $\sigma(v \rightarrow v')$
presents some specific features \citep{neugebauer96}.
The first one consists of two narrow peaks at $v'=\pm v$,  where the one corresponding
to forward scattering ($v'=v)$ is many orders of magnitude higher than
that of backward scattering ($v'=-v$); in this case there is no photon
emission. The second feature, associated with photon emission, is a broad peak around 
the energies $E'_{\rm el}=\pm (E_{\rm el}- \ecyc)$, where $E_{\rm el}'=1/2 \me v'^{2}$ 
and $E_{\rm el}=1/2 \me v^{2}$ are the electron energies after and before scattering, respectively. 
In the latter
case the intermediate state of the Coulomb scattering brings the electron from
the fundamental to higher Landau levels ($n \geq 1$), with a  decay time which is
much shorter than the particle relaxation time.
The outcome is a significant loss of the initial electron energy, which causes
a depopulation of the tail of the Maxwellian distribution that can be described  
by using a simple electron-proton Coulomb interaction.
The energy threshold of this process implies that the higher is the electron temperature $\kte$ 
compared to the magnetic field, the more efficient is the depopulation effect.

Numerical computations performed by R99 showed that the emissivity has a narrow
energy peak at the cyclotron energy for photons emitted in the direction perpendicular to the magnetic field, 
while the peak gets  broader for photons progressively
emitted in the magnetic field direction. When averaging over all the possible
directions, the net result can be \emph{qualitatively} described as a broad cyclotron emission feature
overlapped to a smooth continuum resembling a non-magnetized \bs\ emissivity
spectrum. However, a full self-consistent treatment  is particularly difficult  
and hardly feasible in a run-time code \citep[see][for a solution]{neugebauer96}.
A suitable approximation can be found by splitting the emissivity 
into two different seed photon sources (see also BW07), and this is the approach we follow in our treatment. 
Instead of modeling the cyclotron line emission with a delta function
$\delta(E-E_{\rm cyc})$, we assume a Gaussian profile with a centroid energy of 
$E_{\rm cyc}$ and a variable width $\sigmacyc$. The latter quantity parametrizes all 
the different effects leading to the intrinsic cyclotron line broadening, such as  
thermal broadening, spatial variations of the magnetic field not related to the dipolar structure, and the release of 
photons at all emission angles (this is of particular importance in case of the phase-averaged spectral 
analyses, as those we carry out in this paper; see Sect.~\ref{sec:observation}).
 
The radiative transfer equation (RTE) in the observer frame for sub-relativistic bulk motion 
and arbitrary geometry has been first derived by \citet[][hereafter BP81]{bp81a}.
This particular form of the RTE has then been widely used in the literature
\cite[e.g.][]{ls82, titarchuk1997, farinelli2008}.
On the other hand, \citet[][hereafter PL97]{psaltis97} outlined how the result derived by BP81 was based on
the wrong assumption that the ratio between the zeroth and the second moment
of the specific intensity is equal 1/3, which is strictly true only in the
the fluid frame where the bulk velocity is zero.
The error introduced in the RTE when deriving it in the observer reference frame is at the first
order comparable to the adimensional fluid velocity $\beta$ (see derivation in Appendix \ref{appendix_b}).
Under particular assumptions for the matter velocity profile, the RTE can be solved 
analytically by using the variable separation method for which 
the solutions are determined by the space boundary conditions for 
a bounded medium \citep{sunyaev1980,ls82, titarchuk1997}. 

It is worth noting that analytical solutions of the RTE with bulk motion, even
in the case of \emph{ad hoc} velocity profiles, can be obtained 
if the energy diffusion operator depends only on the electron temperature, assumed
to be constant through the medium.
If second-order bulk Comptonization effects are taken into account, 
a space-dependent term containing  a $\beta^2$ factor appears in the energy operator. In this case,
the variable-separation method can not longer be adopted and solutions
must be obtained through numerical methods.
The contribution of the space dependent term to the solutions of the RTE has been investigated
by, e.g., \cite{psaltis97}; comparison between analytical and numerical solutions
can be found in \cite{titarchuk1997}.  As this term provides in all cases more reliable solutions 
to the RTE, although with an increasing effect depending on the magnitude of $\beta$, we included it in our model.

In the presence of a strong magnetic field ($B \ga 10^{12}$ G), two polarization radiation modes
arise as a consequence of the vacuum polarization \citep[e.g.][]{ventura79}. These two modes are 
commonly labelled as ordinary (O) and extraordinary (E). 
The approximate electron-scattering cross-section for O-photons is 
\begin{equation}
\label{sigma_ord}
\sigma_{\rm ord}(E, \varphi)=\sigma_{\rm T} [{\rm Sin}^2\varphi+ k(E) {\rm Cos}^2\varphi],
\end{equation}
where 
\begin{equation}
   k(E)\equiv  \begin{cases}
    1 & {\rm for}~~ E \geq \ecyc, \\
    (E/\ecyc)^2 & {\rm for}~~ E < \ecyc.   
  \end{cases} 
  \label{k_sigma}
\end{equation} 
The approximate cross-section of the E-photons is instead
\begin{equation}
\label{sigma_ext}
\sigma_{\rm ext}(E, \varphi)=\sigma_{\rm T} k(E)+ \sigma_{\rm l} \Phi_{\rm l},
\end{equation}
where $\Phi_{\rm l}$ is a normalized function of the line profile, and
\begin{equation}
\sigma_{\rm l}=1.9 \times 10^4 \sigma_{\rm T} B^{-1}_{12}.
\end{equation}
In equations (\ref{sigma_ord}) and (\ref{sigma_ext}), $\varphi$ is the
angle between the photon propagation direction and the magnetic field
\citep{Arons1987}.
Both O-photons and E-photons 
interact with the electrons of the plasma with an efficiency dictated by
their relative cross-sections, but mode conversion also occurs at
a given scattering rate \citep{Arons1987}.
A system of two equations for both polarization modes containing the transition-rate
term should thus be solved. 

As outlined by BW07, the angular dependence of
the cross-sections must be replaced by some angle-average values in order to treat the problem in
analytical and numerical codes. Additionally, the explicit
dependence on energy in equation (\ref{k_sigma}) is neglected and instead we use
energy-averaged cross-sections.
On one hand, E-photons mostly escape from the bounded configuration, 
except for energies close to the cyclotron value and its harmonics where the scattering
cross-section becomes highly resonant (see equation [\ref{sigma_ext}]) and
photons are substantially trapped in the line.
On the other hand, O-photons are more efficiently scattered,
particularly at orthogonal propagation angles with respect to the magnetic field
(equation [\ref{sigma_ord}]). 
Building on the general approach of BW07, it has been assumed as an underlying assumption 
in the \compmag\ model that the continuum Comptonization spectrum has to be computed only 
for O-photons, while the contribution of E-photons is phenomenologically taken into account 
by using Gaussian absorption features in the X-ray spectra.

The RTE for the zeroth-moment photon occupation number $n(x)$ in the cylindrically symmetric case 
can be written as 
\begin{equation}
\label{fp_eq}
\begin{split}
\frac{\partial n}{\partial t} - j_{\rm cyc}(x) - j_{\rm bs}(x)   = \\
 -v\frac{\partial n}{\partial z}+\frac{dv}{dz}\frac{x}{3}\frac{\partial n}{\partial x} 
+\frac{\partial}{\partial z}\left(\frac{c}{3 \ne \sigma_\parallel}\frac{\partial n}{\partial z}\right) - \frac{n}{t_{\rm{esc}}}\\
  +  \frac{\ne \overline{\sigma} \Theta c}
{x^2}\frac{\partial}{\partial x}\left[x^4\left(n+f_{\rm b}
 \frac{\partial n}{\partial x}\right)\right] \\
 - \aff (x) n + \beta M \left[\aff (x) + x \frac{\partial \aff(x)}{\partial x}\right] n , \\
\end{split}
\end{equation}
where $x=E/\kte$, $\Theta=\kte/\me c^2$,  $f_{\rm b}=1+ \me v^2/3\kte$,
$\beta=v/c$, $\sigma_\parallel=10^{-3} \sigma_{\rm T}$, $\overline{\sigma}=10^{-1}\sigma_{\rm T}$,
and $\sigma_{\rm T}$ is the Thomson cross-section.
\noindent
Equation (\ref{fp_eq}) can be easily derived from equation (18) of BP81 by adopting a few modifications. 
The first one is the addition of the second-order
bulk Comptonization term $f_{\rm b}$ in the energy distribution operator, as previously
discussed (see equation [33] of PL97 for its derivation).
The second modification is the inclusion of the free-free absorption as expressed in the observer reference frame 
with an accuracy of order of $v/c$ -- higher terms of the series expansion are of order of $v^2/c^2$ and can thus  
be neglected (see equation [B4] of PL97). The free-free absorption was neither considered in the earlier treatment 
of F12, nor in the analytical treatment by BW07. The absorption coefficient can be written as  
\begin{equation}
\label{alpha_ff}
\aff(x)=4.2\times 10^{-9} \frac{ n^2_{19} (e^x -1)  G(x)}{x^3 \Theta^{7/2}},
\end{equation}
where
\begin{equation}
\label{gaunt}
G(x)=e^{-x/2} K_0(x/2)
\end{equation}
is the Gaunt factor and $K_0(x/2)$ is the modified Bessel function of the second kind
\citep{titarchuk88}.

It is important to note that the expression for $\aff$ used in equation (\ref{fp_eq})
is for unmagetized plasma. In the presence of a strong magnetic field, the 
behavior is different and depends on the photon polarization mode: for O-photons
$\aff$ is a smooth function of energy which can be well approximated
by that of unmagnetized case. For E-photons, the Gaunt factor exhibits strong
resonance at the cyclotron fundamental energy and its harmonics $E_{\rm n}=n \ecyc$
\citep{pavlov76}. 
Since equation (\ref{fp_eq}) deals with O-photons, the expression for the
absorption coefficient in equation (\ref{alpha_ff}) is a
good approximation, while the free-free absorption resonances of E-photons are not treated in our continuum model.
To take into account their contributions in the source spectra, we use 
phenomenological Gaussian absorption line components. The E-photons mainly interact with the electrons via the cyclotron resonance, but this is effectively an elastic scattering process that preserves the angle-averaged photon distribution \citep{nagel1980}. Hence no term describing this process should be included in the RTE.

The function $M$ in the last term of equation (\ref{fp_eq}) is the first 
moment occupation number, defined as (see BP81) 
\begin{equation}
M = \frac{1}{3} \left(\frac{\partial n}{\partial \tau} + \beta x \frac{\partial n}{\partial x}\right),
\label{firt_moment}
\end{equation} 
The addition of the  photon escape term is linked to the solution of the RTE 
along the vertical direction. Actually, the system has
a finite size also along the radial direction, and the rate of
photon escaped through the walls of the accretion column must be taken into
account. In principle, one should find solutions of the RTE considering both vertical 
and radial space coordinates \citep{davidson73}.
However, together with the addition of the energy operator, the problem would become 
cumbersome and best suitable to be treated by Montecarlo codes
\citep{odaka2013, odaka2014}. 
We parametrize this effect using 
the characteristic photon time escape defined as $t_{\rm esc}=r_0 \tau_{\perp}/c$, 
where $\tau_{\perp}$ is the optical depth perpendicular to the
magnetic field defined assuming a cross-section $\sigma_{\perp}=\sigma_{\rm T}$.

As previously mentioned, we consider the \bs\ and cyclotron emissions 
as separated sources of seed photons in equation (\ref{fp_eq}) 
(see  Appendix \ref{appendix_a}). It is worth noting that the emission
coefficient expressed in the observer frame for subrelativistic flows 
comprises a zeroth term plus correction factors of the order of $v^2/c^2$ 
(see equation [B4] of PL97). As the accuracy of the equation (\ref{fp_eq}) is 
$\sim v/c$, with the exception of the Comptonization operator, we only retained the 
zeroth-order term for the emission coefficient. 

Using the definition $d\tau=\ne \sigma_\parallel dz$ for the optical depth
along the $z$ axis, performing a logarithmic sampling of the adimensional
energy through the change of variable $x=e^q$, and defining $n=J/x^3$, after
some lengthy calculations, we obtain the adimensional equation for the zero
moment intensity as 
\begin{eqnarray}
\label{fp_eq_b}
\frac{\partial J}{\partial u}-\mathcal S_{\rm cyc}(q, \tau)  - \mathcal S_{\rm BS}(q,\tau)  = 
\left[1+\frac{\me v(\tau)^2}{3\kte}\right]\frac{\partial^2 J}{\partial q^2}\nonumber\\
+\left[e^q-3+\hat{\delta}-\frac{\me v(\tau)^2}{\kte} + \beta \Phi(q)\right]\frac{\partial J}{\partial q}\nonumber\\
+ \left[e^q-3\hat{\delta}-\frac{\xi^2 \beta(\tau)^2}{H} - \Psi (q)  -3 \beta \Phi (q)\right]J+\frac{1}{3H}
\frac{\partial ^2J}{\partial \tau^2}\\\nonumber
-\left[\frac{\beta(\tau)}{H}+ \Phi (q)\right  ] \frac{\partial J}{\partial \tau},
\end{eqnarray}
where $u=t \ne \overline{\sigma} c H$,   $\xi=15.5 r_0/\dot{m}$, and $\hat{\delta}=1/(3 H) dv/d\tau$, with $H=(\sigma/\sigma_\parallel) \Theta$.

\noindent
The parameter $r_0$ contained in the $\xi$ term, is the adimensional column radius expressed in units of NS \scw\ 
radius through the relation $R_0= R^{\rm scw}_{\odot} m r_0$ km, while
$\dot{m}=\dot{M}/\dot{M}_{\rm Edd}$ is the adimensional accretion rate in Eddington units.

\noindent  
The explicit form of the cyclotron seed photon term as a function of the $x$ variable  is
\begin{eqnarray}
\label{source_cyc}
 \mathcal S_{\rm cyc}(x)=1.7 \times 10^{-8} ~n_{19} H(x_{\rm cyc}) e^{-x_{\rm cyc}} x^{-2} B^{-3/2}_{12}\\\nonumber \Theta^{-4} g(x) ,
\end{eqnarray} 
with $x_{\rm cyc}=E_{\rm cyc}/\kte$ and $B_{12}=B/10^{12}$. We also used the normalized Gaussian function 
\begin{equation}
g(x)=\sqrt{\frac{2}{\pi} } \frac{1}{\sigma_{\rm cyc}} e^{-(x-x_{\rm cyc})/2\sigma^2_{\rm cyc}},
\label{gauss_cyc}
\end{equation}
in place of the $\delta$-function corresponding to the monochromatic line emission case.
The \bs\ source term is 
\begin{equation}
\label{source_bs}
 \mathcal S_{\rm bs}(x)=3.8 \times 10^{-13} G(x)~n_{19}~\Theta^{-9/2} e^{-x},
 \end{equation}
with $G(x)$ given in equation (\ref{gaunt}).
The derivation of the source terms in equation (\ref{source_cyc}) and (\ref{source_bs})
is reported in Appendix \ref{appendix_a}.
The functions $\Phi$ and $\Psi$ in equation (\ref{fp_eq_b}) are related to the zeroth
and first order term of the free-free absorption coefficient and are defined
as 
\begin{equation}
\Psi(x)=c_1 \frac{ n_{19}(e^x-1) e^{-x/2}K_0(x/2) }{x^3 \Theta^{9/2}},
\end{equation}
and
\begin{eqnarray}
\Phi(x)=\frac{n_{19}\beta}{3 x^3 \Theta^{9/2}} \left\{K_0(x/2) \left[{\rm c_1}x {\rm Cosh}(x/2)\right. \right] \\\nonumber
\left. -{\rm c_2}{\rm Sinh}(x/2)-  {\rm c_1}~x~K_1(x/2) {\rm Sinh}(x/2) \right\},
\end{eqnarray}
where $c_1=2.1 \times10^{-13}$ and $c_2= 4 c_1$.

Equation (\ref{fp_eq_b}) is solved by using the numerical procedure described
in F12. We considered as natural boundary conditions that: (i) the intensity 
vanishes at the extremes of the energy domain and at the top of the column; (ii) 
the approximation reported in equation (31) of F12 holds at the base of the accretion column. 
We assumed that the NS magnetic field within the accretion column is dipolar, and thus $B_{12}(z)=B^{\rm ns}_{12} (z/z_0)^{-3}$,
where $B^{\rm ns}_{12}$ is the magnetic field strength at the base of the accretion column (i.e., the NS surface at $z_0$). 
In the updated version of the \compmag\ model being developed here, the free parameters regulating the cyclotron emissions 
to be determined through the fit to the X-ray data are thus $B^{\rm ns}_{12}$ and $\sigma_{\rm cyc}$. 
The additional improvement with respect to the former \compmag\ version that we introduce here is  
the possibility to treat the height of the accretion column $\zmax$ as a free parameter.
We also stress that in the new version of the model the mass accretion rate has been used as a free parameter 
in \xspec\ in place of the vertical optical depth adopted before (see also Table \ref{tab_newcompmag}).  
The relation between these two quantities depends on the form of the velocity profile of the accreting material. 
Starting from the continuity equation 
\begin{equation}
\label{cont_equation}
\dot{M}=\pi R^2_0 m_{\rm p} \ne \left|v_{\rm z}\right|,
\end{equation}
and defining the first adimensional velocity profile as
\begin{equation}
\beta=-\beta_0 \left(\frac{z_0}{z}\right)^\eta,
\label{vel_1}
\end{equation}
one obtains that the optical depth measured at the top of the column is
\begin{equation}
\label{tau_vel1}
\tau_0= 2.17 \times 10^{-3}\frac{\dot{m}(\zmax^{1+\eta}- z_0^{1+\eta})}{ \mathscr{A} r_0^2 (1+\eta)}, 
\end{equation}
where $\mathscr{A}=\beta_0 z_0^\eta$ while $\beta_0$ is the matter
velocity in units of $c$ at the NS surface, and $z_0$ is the adimensional NS radius. 
Given the definition of $r_0$, we obtain $H= R^{\rm scw}_{\odot} m z_0$ km.
Here and throughout the paper we assumed $m=1.4$ and $R_{\rm ns}=10$ km, implying
$z_0=2.42$.

If the inflowing matter is assumed to decelerate towards the NS surface according to the 
law (see also equation [27] of BW07)
\begin{equation}
\beta=- \frac{10.4 r_0}{\dot{m} z_0} \tau,
\label{vel_2}
\end{equation} 
then, the relations between $\tau_0$ and $\dot{m}$ is given by  
\begin{equation}
\label{tau_vel2}
\tau_0=\frac{\dot{m} z_0^{1/2} (\zmax- z_0)^{1/2}}{50 r_0^{3/2}}.
\end{equation} 
It is worth pointing out that the adimensional velocity of equation (\ref{vel_2}) is
limited to be less than unity at the top of the column.
In the above equation, $\tau_0$ is the approximated vertical optical depth 
calculated with the reduced Thomson cross-section for photons propagating
along the lines of the magnetic field:
\begin{equation}
\tau_0 = \int_{z_0}^{z_{\rm max}} \ne \sigma_\parallel \mathrm{dz'}.
\end{equation}

The above reported equations are implicitly derived under the assumption
that the accreting column has a pure cylindrical shape with constant
radius over height.
Actually, for a dipolar magnetic field the column has a cone-like shape
of half-angle $\theta_{\rm c}$ given by \citep{frank2002}

\begin{equation}
{\rm Sin^2} \theta_{\rm c}= \frac{R_{*}}{R_{\rm m}} {\rm Sin^2} \psi,
\end{equation}

where $R_{*}$ is the NS radius, $R_{\rm m}$ is the radius at which matter
starts to be channeled by the magnetic field lines, and $\psi$ is the
inclination angle between the magnetic axis and the NS equatorial rotating
plane.

Finally, the last important improvement we introduced in the new version of the \compmag\ model 
is related to the geometry of the X-ray emission. We now differentiate between the pencil beam case, 
when the X-ray radiation is emitted upwards with respect to the NS surface, and the fan beam geometry, 
when the radiation is released from the lateral boundaries of the accretion column. 
Assuming a cylindrical accretion column, the adimensional photon flux in the pencil beam case 
is given by 
\begin{equation}
F^{\rm p}(x) = \frac{4 \pi x^3}{3} \left(\frac{\partial n}{\partial \tau} + \beta^{\rm t} x \frac{\partial n}{\partial x}\right),
\label{flux_geom1_adim}
\end{equation}
where $\beta^{\rm t}$ is the adimensional accreting matter velocity at the top of
the column.
 
In order to pass to physical units, it is necessary to consider the relation between the
intensity and the occupation number 
\begin{eqnarray}
\label{intensity_occnumb}
J(E) \approx 3 \times 10^{31} (\kte/{\rm keV})^3 x^3 n(x)\\\nonumber
 ~{\rm keV cm^{-2} s^{-1} keV^{-1} ster^{-1} },
\end{eqnarray}
which can be easily derived once defining $h\nu= x \kte$ (see equation [\ref{occ_numb_a}]). 
The \compmag\ normalization $\ncomp$ is defined in this case as
the ratio between the energy flux multiplied by the top-column
surface $S=\pi R_0^2$, and divided by 4$\pi D^2$, where $D$ is the source
distance. Using equations~(\ref{flux_geom1_adim}) and (\ref{intensity_occnumb}), we obtain 
\begin{equation}
\mathscr{F}^{\rm p}(x) \approx 0.02 (\kte/{\rm keV})^3 ~ \frac{r_0^2}{d^2_{10}} F^{\rm p}(x)  ~{\rm keV~ cm^{-2} ~s^{-1} ~keV^{-1}},
\label{flux_geom1}
\end{equation}
where $r_0$ and $d_{10}$ are the adimensional column radius and source distance
in units of 10 kpc, respectively.

In the fan-beam case, the high optical depth allows the photons to 
diffuse and escape through the lateral boundaries of the accretion columns.
As \compmag\ is a one-dimensional model and the variations of the
physical quantities can be computed only along the vertical ($z$) and not the radial direction, 
we adopted some simplification. More specifically, considering the  zeroth-moment specific 
intensity $J(x,\tau)= x^3 n(x,\tau)$ at the walls of the column and assuming a uniform 
brightness $B(x, \tau)=\pi J(x,\tau)$,  the vertically-integrated adimensional photon 
flux is given by 
\begin{equation}
\label{flux_geom2_a}
F^{\rm b}(x)= \frac{{\rm P_{cyl}}}{4 \pi D^2} \int_{0}^{\tau_0}  B(x, \tau) \frac{dz}{d\tau} d\tau,
\end{equation}
where $dz/d\tau$ can be easily derived by differentiating either equation (\ref{tau_vel1})
or equation (\ref{tau_vel2}), and ${\rm P_{cyl}}=2 \pi R_0$ is the column perimeter.

From equation (\ref{intensity_occnumb}) and (\ref{flux_geom2_a}) we then obtain
\begin{eqnarray}
\label{flux_geom2_b}
\mathscr{F}^{\rm b} (x)\approx 9 \times 10^{-3} (\kte/{\rm keV})^3~\frac{r_0}{d^2_{10}} \int_{0}^{\tau_0} B(x,\tau) \frac{dz}{d\tau} d\tau \\\nonumber
~{\rm keV~ cm^{-2} ~s^{-1} ~keV^{-1}}.
\end{eqnarray} 
Note that we expressed the fluxes in equations (\ref{flux_geom1}) and (\ref{flux_geom2_b})
in keV because these are the standard units when writing \xspec\ models.
The conversion to cgs units is performed by the fitting package at the end of the runs.

The \compmag\ normalization factor $\ncomp$, as derived in equations (\ref{flux_geom1}) and (\ref{flux_geom2_b}), 
deserves however some considerations. Once the source distance is fixed, the other parameters
that determine the observed flux are $\kte$ and $r_0$, to be inferred from the fit to the X-ray data.  
This means that, in principle, $\ncomp$ should be kept \emph{fixed} during
the fit to a value of the order of $\sim 1/d^2_{10}$, where $d_{10}$ is the distance of the source 
in units of 10\,kpc. 
However, in the computation presented above (and also reported in BW07) we considered 
only the approximated case of a purely cylindrical accretion column and a flat space-time. 
In a more realistic situation, it is likely that hydrodynamical and General Relativity effects 
will affect the spectral energy distribution of the X-ray radiation. 
In particular, the gravitational-lensing effect, due to the 
curved space-time close to the NS surface, is expected to enhance the emerging X-ray flux at some angles
compared to the Newtonian case. For this reason, the normalization $\ncomp$ should be 
proportional to $f$$\times$$1/d^2_{10}$, where $f > 1$ is a numerical factor that 
parametrizes all the effects that can not be taken (yet) into account in the present version of the 
\compmag\ model.

\subsection{New convergence procedure of the algorithm}

In the first version of the \compmag\ model proposed by F12, the authors developed an empirical 
criterion to halt the iteration during the resolution of the radiative transfer equation in 
the pulsar accretion column. The code first estimated the approximated energy index $\alpha$ 
from the fit to the source spectrum in the energy range $E_{\rm min} < E < E_{\rm max}$, where 
$E_{\rm min} > 3 \ktbb$ and $E_{\rm max} \la 10$ keV. These boundaries were chosen in order 
to limit the fit to the region where the energy spectral distribution is better approximated with 
a power-law, and the iteration stopped when $1-\alpha_{\rm m}/\alpha_{\rm m+1} < \varepsilon$.
Additionally, the vertical optical depth $\tau$ was always divided into $N=10$ bins independently 
of its value to achieve a reasonably fast computing time. 

This method was shown to be well suited to perform fits to the relatively low quality data that 
were available to F12. The data that will be used in the present paper to study some among the brightest 
X-ray pulsars in our Galaxy (see Sect.~\ref{sec:observation}) require, instead, a significant improvement 
of the convergence procedure for the resolution of 
the radiative transfer problem in the accretion column. We verified that the presence of cyclotron emission 
features in the spectra of these sources makes the estimation of the spectral slope through a two-point linear interpolation 
largely uncertain, resulting in the non-convergence of the algorithm or an unsustainable high number of numerical iterations. 
We thus changed the initial differential approach of F12 to the convergence problem into an integral one. 
Defining $u^m_{i,j}$ and $u^{m+1}_{i,j}$ as the computed values of the source energy spectral distribution at 
the $m^{\rm th}$ and $m+1^{\rm th}$ iterations, with $(i,j)$ identifying the steps in the pre-defined grid of energy and optical 
depth, the new stopping criterion for the convergence of the algorithm can be expressed as 
\begin{equation}
| u^{m+1}_{i,j}-u^{m}_{i,j} | < \varepsilon ~~~~\forall i,l.
\end{equation} 
The maximum allowed value of $\varepsilon$ is fixed by verifying at each iteration step that 
the \chiq\ obtained from the fit to the data is monotonically decreasing. We checked {\it a posteriori} 
that $\varepsilon=10^{-4}$ allows to achieve the required accuracy in the estimate of the model 
parameters during all fits performed, maintaining at the same time a reasonably short computational time. 

We also checked that for the new dataset used in this paper different values of the number $N_{\tau}$ 
of bins selected for $\tau$ resulted in remarkably different values of the model parameters obtained from the fits. 
After a number of trials, we found that fixing $N_{\tau}=50$ provides an optimal solution, as higher values do not 
significantly alter the results of the fits and the computational time of each fit is still reasonably short 
(a few hours on a quad-core Linux machine with a 2.4 GHz processor).

\section{The sources}
\label{sec:sources}

\subsection{\cen}

The high-mass X-ray binary \cen
was discovered by \emph{Uhuru} \citep{Giacconi1971} and is known to host 
a $\sim$4.8\,s spinning NS with an estimated mass of 1.34$\pm$0.15\,$M_\odot$ \citep{VanderMeer2007}, and 
a O 6-8 III supergiant star with a mass of 20.5$\pm$0.7\,$M_\odot$. The orbital period of the system is 
$\sim$2.1 days, and the NS is on a nearly circular orbit eclipsed by the massive companion for 
about 20\% of the time \citep{Hutchings1979,Ash1999}. 
The distance to the source is poorly constrained. A lower limit of 6.2\,kpc was suggested by \citet{Krezminski1974}, 
while \citet{Thompson2009} derived more recently an estimate of 5.7$\pm$1.5\,kpc. For the purpose of our work
and to ease comparisons with published works, we adopt hereafter a distance of 8\,kpc. 
The system is known to undergo episodes of both disk and wind accretion, as shown by the positive and negative 
fluctuating derivative of the NS spin period \citep[see e.g.,][]{bildsten1997}. 

Prominent fluorescence Iron lines for different ionization states are present 
at 6.4, 6.7, and 6.97~keV in the X-ray spectrum of the source and have been 
reported at all orbital phases with variable intensities.
Observations performed with the gratings on-board Chandra also evidenced 
the presence of Doppler broadened \ion{Si}{XIII}, \ion{Si}{XIV}, and \ion{Fe}{XXV} complexes caused by 
recombination in a photoionized plasma \citep{Wojdowski2003,iaria2005}. It has been argued that the fluorescent 
lines originate relatively close to the NS (probably in the outer regions of the accretion disk), while the other lines 
are produced at larger distances from the NS in the wind of the massive companion \citep{iaria2005,naik2011}. 
The stellar wind velocity measured from these latter lines is significantly smaller than that expected for a 
O-type star, providing convincing indications for a strong perturbation of the wind by the NS X-ray radiation.

The source X-ray spectrum is generally well described by a power-law model with exponential cutoff and a 
Gaussian-like emission feature centred at $\sim$13\,keV \citep{suchy2008}. 
A cyclotron absorption feature was detected at $\sim$30\,keV \citep{burderi2000}.  
A partial covering component has also been introduced in some cases to model the variable column density of the 
source and interpret the hour-long dips in the source lightcurve \citep[see, e.g.,][]{naik2011}.
The X-ray variability of \cen\ ranges from time-scales comparable to the orbital period, down to 
fractions of the NS spin period. The spectral energy distribution also shows a remarkable dependence on 
the spin and orbital phase \citep[see, e.g.,][and references therein]{suchy2008, Raichur2008, naik2011}. 

\subsection{\fu}

Hard X-ray radiation from \fu\ was discovered first 
in 1969 \citep{johns1978}, during one of the giant type II outbursts displayed by the source. 
In that occasion, the peak luminosity of the event reached a few $10^{37}$\,erg/s, two orders of 
magnitude above the usual quiescent emission level of the source. A 
periodicity of 3 years for the outbursts from \fu\ was later suggested by \citet{whitlock1989}. 
The source orbital period was measured by \citet{rappaport1978} using \textsl{SAS} data at 
$P_\mathrm{orb}=24.3\,\mathrm{d}$. These authors also estimated the eccentricity of the orbit 
as $e=0.34$, and its semimajor axis $a_X \sin i = 140.1\,\mathrm{lt-s}$. 
X-ray pulsations with a period of $P_S=3.6$\,s were discovered by 
\citet{cominski1978}. 

The source has been widely studied in the optical and IR bands,
leading to the identification of its optical counterpart 
\citep[V~635~Cas;][]{johns1978} and the determination of its distance 
\citep[7-8\,kpc;][]{negueruela2001a}. 
The X-ray spectrum of the source above $\sim 10$\,keV is usually well 
described by power-law model with an exponential cut-off \citep[e.g.][]{coburn2002}. 
The centroid energy of the fundamental cyclotron line is at $\sim$11~keV, and up to six 
harmonics have been observed and reported in the literature 
\citep[][F09]{wheaton1979,white1983,heindl1999,santangelo1999}. 
This is a peculiar characteristics of \fu,\ as in all other sources displaying cyclotron lines 
only the second harmonic is usually detected \citep[the only exception being V~0332+53, which also 
showed evidences for the third cyclotron line harmonic;][]{coburn2002, orlandini2004, tsygankov2006, isabel2007}. 
The variations of the centroid energy of the fundamental cyclotron line with the source luminosity 
were discussed by \citet{mihara2004}, \citet{nakajima2006} and \citet{russi2007}, but these results 
were later criticized by \citet{mueller2013}. The latter authors showed that slightly different fits to the continuum 
emission could produce artificial (but significant) variations of the centroid energy. 
On the other hand, F09 presented the results of the fit to the broad-band X-ray spectrum of \fu\ with the 
BW07 model, aiming at constraining the properties of the X-ray emitting region and the high energy 
radiation mechanisms. These authors showed that the cyclotron emission has a dominant role in cooling the 
material that is flowing through the accretion column in this source, and inferred a magnetic field 
strength from the cyclotron continuum that is approximately a factor of 2 lower than the one estimated 
from the centroid energy of the fundamental cyclotron line. This suggested that the heights of the continuum 
and cyclotron line formation along the pulsar accretion column were substantially different.

\subsection{\her}

\her\ hosts a NS with a spin period of 1.24\,s and a mass of 1.5$\pm$0.3 M$_\odot$, 
orbiting around an A/F donor star, HZ~Her. 
The orbital period of the system is 1.7~d and the high inclination angle ($\sim$85~degrees) 
with respect to the observer line of sight gives rise to extended X-ray eclipses, lasting about 
20\% of the binary orbit \citep{Scott2000}. The super-orbital modulation of the source X-ray 
emission with a period of 34~d is usually ascribed to the presence of a warped accretion disk 
\citep[][and references therein]{staubert2009a} or to the free precession of the NS-disc system 
\citep{Staubert2013}. During the super-orbital period,  \her\ displays a ``main-on''  
state lasting 10-11~days, during which the source achieve the highest X-ray luminosity, and 
a short slightly dimmer state which typically last 5 to 7~days. At other super-orbital 
phases, the source is still visible in the X-ray domain, although at a much lower luminosity 
level. 

\her\ is characterized by a low Galactic absorption and is situated at a distance of 
6.6$\pm$0.4\,kpc \citep{Reynolds1997}. The source X-ray spectrum shows emission lines produced from the 
photoionized accretion disc and on the companion surface, as shown by 
Chandra and XMM-Newton grating observations \citep[see, e.g.,][and references therein]{Zane2004,Jimenez2005,Ji2009}.
The fundamental CRSF was initially detected with a centroid energy of 40\,keV \citep{Truemper1977}, but it has been 
shown that such energy decreased in the past 20 years by 4.2~keV \citep{Staubert2014,klochkov2015}. 
The centroid energy is also known to positively correlate with the overall source X-ray flux over a wide range of different 
timescales \citep[from seconds to days;][]{staubert2007,Klochkov2011}. This behaviour has been interpreted in terms 
of the so-called sub-critical accretion regime, during which the height of the cyclotron scattering forming region 
decreases at higher mass accretion rates \citep{Becker2012}. This conclusion was criticized by \citet{Mushtukov2015}. 
These authors showed that, at the luminosity of \her\ where the cyclotron line is typically monitored, the source should be 
in the super-critical accretion regime, and thus an opposite trend of the cyclotron centroid energy with the source luminosity should 
be observed. An alternative possible solution to this problem is that the magnetic field of \her\ is strongly non-dipolar, 
resulting in multiple accretion sites on the NS surface with non-cylindrical sections \citep{Shakura1991}. It is worth noticing that a 
the combination of a pencil and a fan-beamed emission from a misaligned dipole would also be able to reproduce the pulse profiles 
of \her\ with a reasonable accuracy \citep{Leahy2004}. The latter conclusion was also independently confirmed by \citet{blum2000}, 
who used a pulse profile deconvolution method to infer the pattern of the X-ray emission close to the NS surface and
the geometrical properties of the corresponding emitting region. 

More detailed studies of the source pulse profile were presented by \citet{vasco2013} and \citet{Enoto2008}. The first authors 
analyzed RXTE observations of \her\ and found a puzzling double peak in the variation of 
cyclotron line energy and photon index before and in correspondence of the source pulse maximum.
\citet{Enoto2008} analyzed a 60\,ks-long Suzaku observation of the source and reported on the possible 
detection of the second cyclotron harmonic close to the maximum of the pulse profile. 
A extensive analysis of three combined Suzaku/NuSTAR observations was carried out by \citet[][hereafter F13]{Fuerst2013},
who used a phenomenological model to describe the source emission during the main-on state.

\section{Observations and data analysis}
\label{sec:observation}

In order to properly test the improved version of the \compmag\ model presented 
in this paper, we made use of a set of high-statistical quality data, providing a broad 
and continuous coverage in the energy range 0.5-100~keV.  
Our sample exploits archive BeppoSAX \citep[see ][
for a description of the mission]{sax} data of \cen and \fu, together with a combined NuSTAR 
\citep{nustar} and Suzaku \citep{Suzaku} observation of \her\ during one of the source main-on state 
(see Sect.~\ref{sec:sources}). We summarize all characteristics of our data-set in Table~\ref{tab:observations}, 
together with the references to previously published papers that made use of these data.  
Note that the two \cen\ observations reported in Table~\ref{tab:observations} were obtained with the source 
at different luminosities and we removed the time intervals corresponding to the source eclipse before carrying out 
any further analysis.  
\begin{table*}[!th]
\caption{Log of the observations of the sources analysed in this paper.}
\begin{center}
 \begin{tabular}{ l c c c c c c }
\hline
\hline
Source\tablefootmark     & Start Time [UT] & Stop Time [UT] & \multicolumn{4}{c}{Exposures [ks]}  \\
             &                   &                   &{\small \lecs}&{\small\mecs}&{\small\hp}&{\small\pds } \\ 

\hline
\cen   (LL)\tablefootmark{a}   & 1996-08-14 12:38 & 1996-08-14 23:46 &  4.3  & 24.5 &  11.2 & 5.5 \\
\cen   (HL)\tablefootmark{b}    &  1997-02-27 19:45   & 1997-02-28 11:00  & -     & 17.9 & 8.9 & 8.2 \\
\fu (Obs. I)\tablefootmark{d} &  1999-03-19 17:05 & 1999-03-20 08:42 & 3.3 & 31.2 & 32.0 & 30.0  \\
\fu (Obs. II)\tablefootmark{e} &  1999-03-26 17:31 & 1999-03-27 17:34 & 5.1 & 53.7 & 42.5 & 48.3 \\
\hline
             &                   &                   & {\small XIS}  &{\small HXD} & {\small FPMA} & {\small FPMB} \\ 
\hline
\her\tablefootmark{c}       &  2012-09-22 04:39  & 2012-09-22 18:33  & 22.6 & 18.7 & 21.9 & 22.1 \\        
\hline
\end{tabular}
\tablefoot{
\tablefoottext{a}{Unpublished \sax data.}\\
\tablefoottext{b}{The same data set was studied by \citet{burderi2000}. As for the present paper, these authors excluded the time 
intervals corresponding to the source eclipse before performing any analysis. \lecs data were not available for this observation  
due to telemetry problems.}\\
\tablefoottext{c}{An analysis of these data is reported in \citet{Fuerst2013}.}\\
\tablefoottext{d}{A study of the source hard X-ray spectrum extracted from this observation is reported by \citet{santangelo1999}.}\\
\tablefoottext{e}{A study of the broad-band spectrum, as derived from these data, is reported by F09.}\\
}
\label{tab:observations}
\end{center}
\end{table*}

Concerning the BeppoSAX observations, we analyzed data from the high energy narrow field instruments
\lecs \citep[0.7-4.0\,keV,][]{lecs}, \mecs \citep[1.5-10.5\,keV,][]{mecs}, \hp \citep[7-44\,keV,][]{hp} 
and \pds \citep[15-100\,keV,][]{pds} after performing the full processing through the standard pipeline
(\texttt{saxdas} v.2.1, running only on heasoft v. 5.2). 
Background subtraction was performed by using the
earth occultation for the \hp, the off-source pointing for the \pds, and the
standard calibration files for the \mecs and the \lecs. The \textsl{LECS} and
\textsl{MECS} source extraction radii were set to $8^\prime$ to optimize 
the signal to noise ratio.
 
We adopted standard reduction procedures for the Suzaku and NuSTAR data, devoting a 
particular attention in evaluating the effect of pile-up during the Suzaku/XIS observation. 
As in the latter case only the outer portion of the instrument chip was active, we made use  
of data exclusively collected through the optimally calibrated XIS3 \citep[see][for further details]{Fuerst2013}. 
Considering the scope of this paper, we limited our analysis to the 3-8.5 keV energy range 
for the Suzaku/XIS, 16-80 keV for the Suzaku/HXD, and 3.5-65 keV energy range for NuSTAR. 
Any spectral contribution coming from the accretion disk and 
the surrounding corona was thus excluded from further analysis. 
We only considered for each source spectra averaged over the entire spin period and did not 
perform a phase-resolved spectral analysis. 

All spectral analyses were performed with the \XSPECB\ version 12.8 \citep{xspec}, using 
the \compmag\ model described in Sect.~\ref{sec:new_compmag}. 
Inter-calibration constants were included in the fits and found to be compatible with typically  
expected values \footnote{See https://heasarc.gsfc.nasa.gov/docs/sax/abc/saxabc/saxabc.html.}. 
The fundamental and all detected harmonics of the CRSF features in each of the considered 
source were accounted for in the fits by using the {\sc gabs} model in \XSPECB:
\begin{equation}
G(E) = \exp\left( - \frac{\tau_n}{\sigma_n\sqrt{2\pi}}e^{-\frac{1}{2}{\left(\frac{E-
\ecyc}{\sigma_n}\right)}^2}\right) \ .
\label{eq:gaus}
\end{equation} 
Here $\tau_n$, $E_{cn}$, and $\sigma_n$ denote the optical depth, energy, and
standard deviation of the absorption for the $n$-th harmonic, respectively. 
The line width is fixed to previously measured values whenever it cannot be robustly 
determined by the fit. For \cen\ and \her,\ we have linked the intensity of the magnetic field in 
the \compmag\ model used to reproduce the continuum to the centroid energy of the fundamental CRSF;
for \fu\ it was linked to the first harmonic, as this is known to be less affected by biases in 
the continuum modelling of the source \citep{mueller2013}.
\begin{figure}
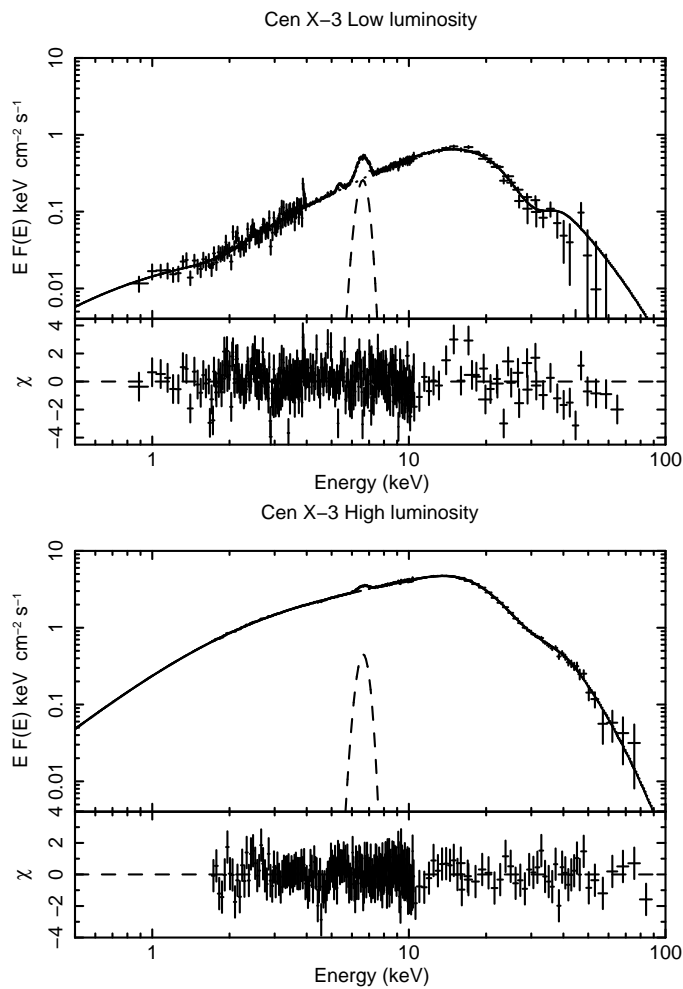

\includegraphics[width=6.5cm, angle=270]{fig1.ps}
\includegraphics[width=6.5cm, angle=270]{fig2.ps}
\caption{Unfolded spectra of the two \sax\ observations of \cen. We also show in the figure the best fit 
\compmag\ model and the residuals from the fit in units of $\sigma$ (bottom panel). All best-fit parameters determined from the fit 
are reported in Table \ref{tab_newcompmag}.}
\label{eeuf_spectra_cenx3}
\end{figure}
\begin{figure}
\includegraphics[width=6.5cm, angle=270]{fig3.ps}
\includegraphics[width=6.5cm, angle=270]{fig4.ps}
\caption{Same as Fig. \ref{eeuf_spectra_cenx3} but for the two \sax\ observations of \fu.}
\label{eeuf_spectra_4u}
\end{figure}
\begin{figure}
\includegraphics[width=6.2cm, angle=270]{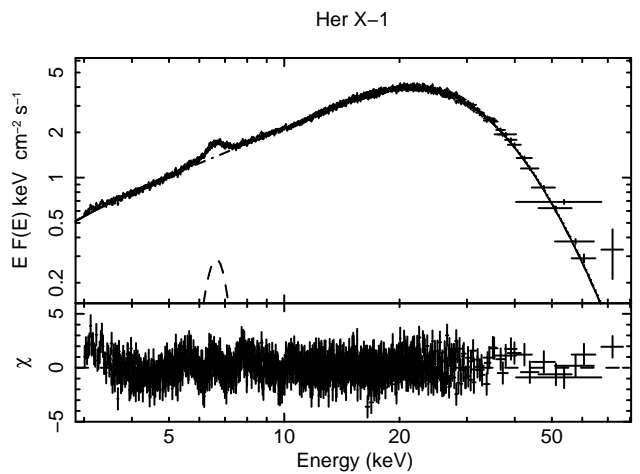}
\caption{Same as Fig. \ref{eeuf_spectra_cenx3} but for the combined \suzaku\ and \nustar\ observation of \her.}
\label{eeuf_spectra_her}
\end{figure}
\begin{figure}
\includegraphics[width=6.5cm, angle=270]{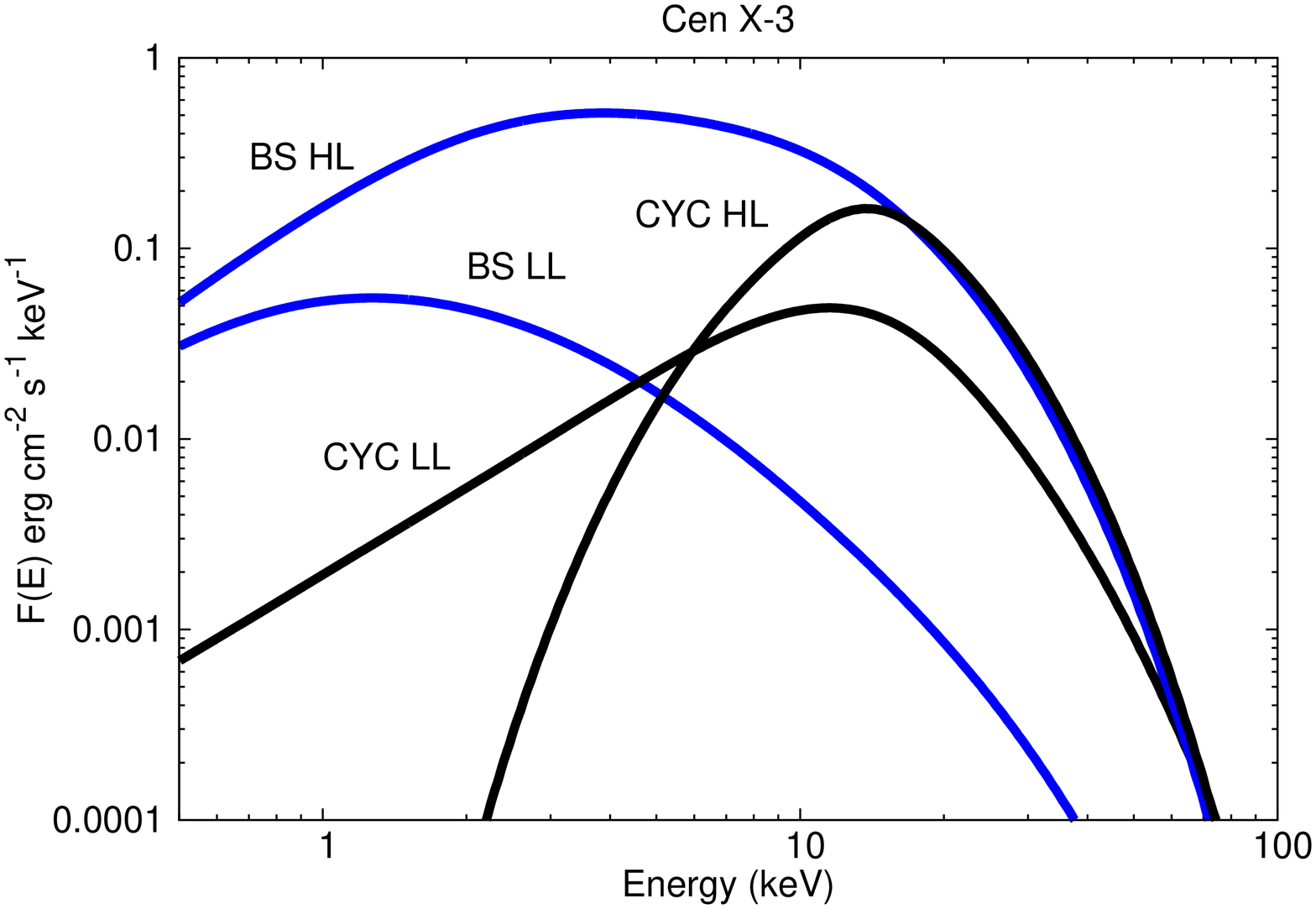}
\includegraphics[width=6.2cm, angle=270]{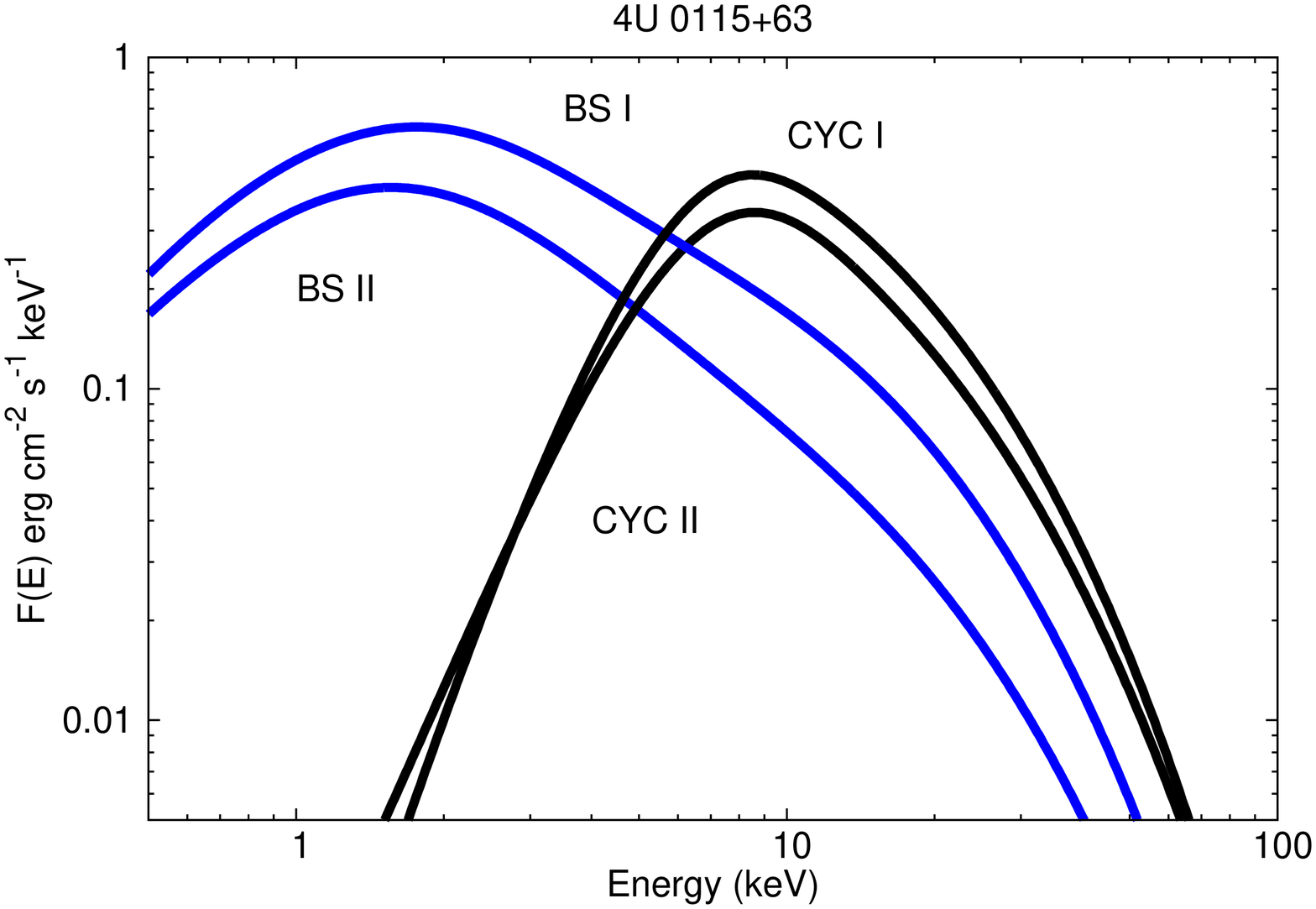}
\includegraphics[width=6.5cm, angle=270]{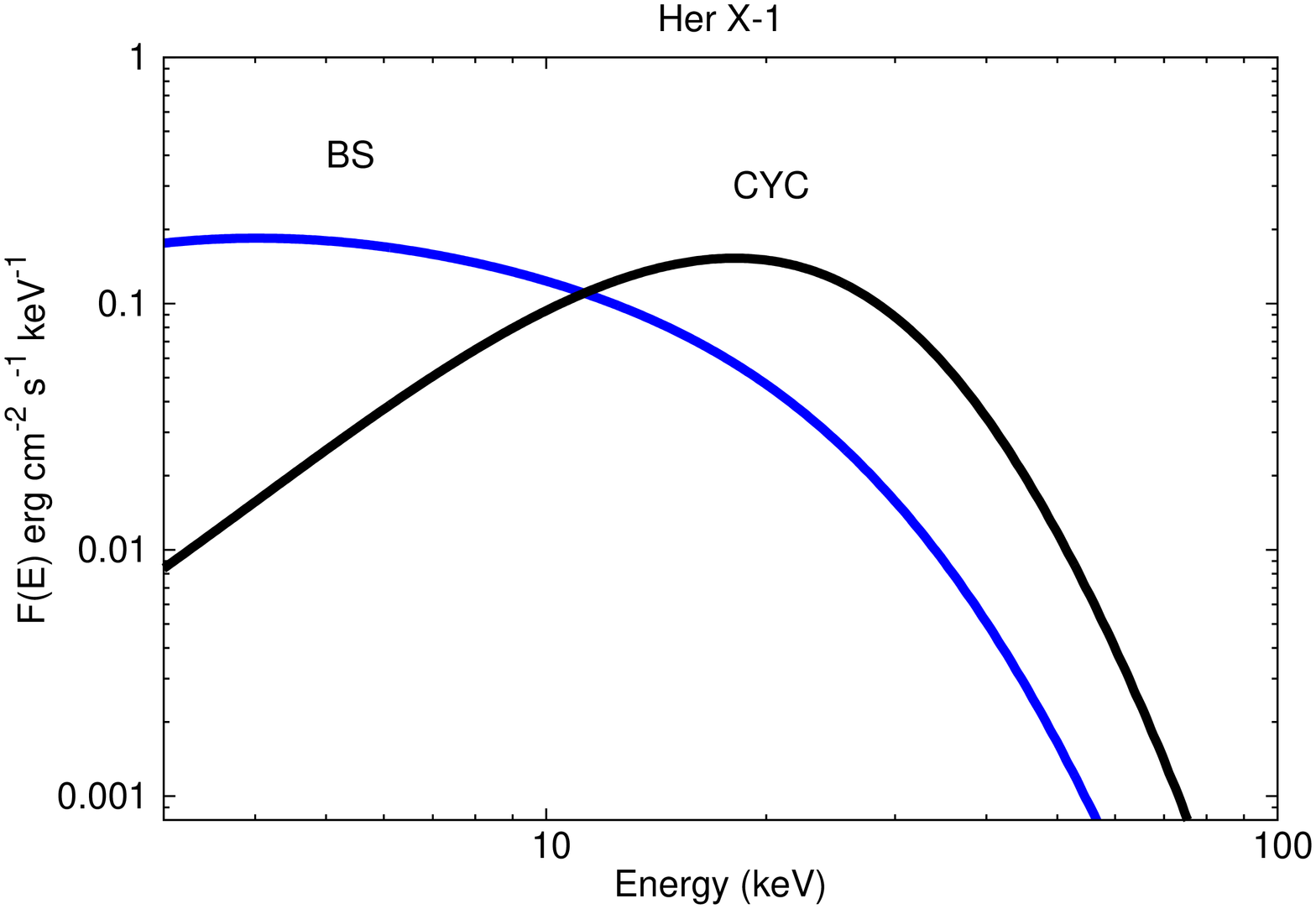}
\caption{From top to bottom: Comptonized \bs\ and cyclotron spectra of \cen, \fu, and \her\ obtained 
from the best-fit models reported in Table \ref{tab_newcompmag}.}
\label{spectra_components}
\end{figure}

\section{Results}
\label{sec:results}

The updated version of the \compmag\ model that we introduced in the present paper 
permits to compute the spectra of X-ray pulsars emerging in both a pencil-beam or a fan-beam 
geometry. The former case can be obtained by using equation~(\ref{vel_1}) for
the velocity profile of the infalling material in the pulsar accretion column and equation~
(\ref{flux_geom1}) to describe the emerging X-ray flux. The parameter $\beta_0$ defines in this geometry  
the terminal velocity of the infalling material at the NS surface. 
The corresponding equations for the fan-beam geometry are (\ref{vel_2}) and (\ref{flux_geom2_b}). 

A preliminary fit to the available spectra considered in this work revealed that 
reasonably good values of the \chiq\ could be obtained with all the 
\compmag\  parameters left free to vary, but the uncertainties on these parameters 
were not very constraining. We verified that this is largely due to the computation in the model of the 
matter velocity profile inside the accretion column, as the latter determines 
the optical depth of the column in the vertical direction and thus in turns the 
total source luminosity. As all sources in this paper are characterized by a 
luminosity that is far above the value at which the fan-beam emission geometry 
is expected to set-in, we first fixed the normalization of the \compmag\ model 
($\ncomp$) and concentrate on the fan-beam case with efficient matter deceleration 
along the accretion column due to the strong radiation pressure (see 
equations (\ref{vel_2}) and (\ref{flux_geom2_b})). This eases 
the comparison with both the previous theoretical
treatment of the model reported in BW07 and its first application discussed by F09. 
For completeness, however, we also provide in the next sessions a discussion about the results 
of the fits to all data with the \compmag\ model and the pencil-beam geometry.

In all cases, we modelled the effect of the interstellar photoelectric absorption on the 
X-ray emission from each of the considered sources with the \phabs\ spectral component in \xspec,\ 
assuming Solar abundances.

\subsection{\cen}
\label{sect:cen}

We first fit the spectra of \cen\ obtained from the two available BeppoSAX observations 
(LL and HL; see Table~\ref{tab:observations}) with the \compmag\ model in the fan-beam geometry case, plus a 
cyclotron Gaussian absorption line centered at $\sim$30~keV. 
The results of the fits were not yet acceptable (\chiq/dof=4049/318 and \chiq/dof=675/241 
for the LL and HL case, respectively) due to evident residuals around 6.7 keV. 
The addition of a Gaussian emission line at this energy significantly improved the results 
of both the LL (\chiq/dof=509/315) and HL (\chiq/dof=167/238) observations.

In the case of the LL observation, some residuals were still visible around 5~keV and below 
2~keV. The former are most likely due to a known feature of the MECS instrument and could be 
modelled by adding a Gaussian component with a fixed width of $\sigma=0.1$ keV. This improvement  
led to \chiq/dof=447/313. The residuals below $\sim$2~keV observed in the LECS data (not 
available during the HL observation) were interpreted them in terms of a second region of 
absorption partially covering the source. This possibility has been already investigated by \citet{naik2011}, 
during the phase-resolved spectroscopy of the source performed on some \suzaku\ data.
The authors found a strongly-variable local absorption component with
$\Nh$ ranging from $\sim 2 \times 10^{22}$ to $\sim 1.3 \times 10^{23}$ cm$^{-2}$, and a 
covering fraction in the range $0.3-1$.
Following the above mentioned work, we added a multiplicative partial covering absorption component (\pcfabs) to our \xspec\ 
model and obtained a significantly improved result (\chiq/dof=369/311) with 
$\Nh \sim 5 \times 10^{22}$ cm$^{-2}$ and a covering fraction of $\sim 0.4$.
In the HL spectrum, this component
was not necessary to obtain an acceptable fit (see Fig. \ref{eeuf_spectra_cenx3}),
possibly due to a higher ionization of the absorbing medium and/or by
a weaker constrain of the absorption in the data set, due to the lack of \lecs\ data.

Although the above discussed model provided a reasonably good fit to both the available  
observations of \cen, we obtained from the \compmag\ best fit parameters a similar accretion rate in both observations. 
This result does not seem to be consistent with the eight times higher X-ray luminosity of the second observation 
(see Table~\ref{tab_newcompmag}). We verified that this computational issue is due due to the degeneracy in the fits 
between the \compmag\ normalization $\ncomp$ and the mass accretion rate $\mdot$, as they both 
regulate the source luminosity (see Sect.~\ref{sec:new_compmag}). 
To avoid this degeneracy, we performed a new fit to the LL data with the \compmag\ 
normalization fixed to the value obtained from the fit to the HL data ($\ncomp \sim 60$).  
The fit with this model provided \chiq/dof=449/312, a slightly 
worst value  than that obtained with a free to vary normalization, but with the accretion rate parameters
$\dot{m}$ now tracing reasonably well the source luminosity variation between the
two observations (see Table~\ref{tab_newcompmag}). No significant changes of the $\Nh$ and the covering fraction 
parameters in the \pcfabs\ components were altered by fixing the \compmag\ normalization.  
It is noteworthy that the determined value of the $\Nh$ is comparable to that measured by \cite{naik2011} during 
their phase-averaged spectral analysis (see their Table~1). The most significant change was recorded for the mass accretion rate estimated 
from the fit, that is now in rough agreement with that expected from the ratio of the X-ray luminosities measured from the 
LL and HL observations (see Table \ref{tab_newcompmag}). For completeness, we also tried to fix the normalization 
of the \compmag\ model in the fit of the HL data to the value obtained from the original fit to the LL data 
(with $\ncomp$ left free to vary). However, in this case, we could not obtain an acceptable fit 
to the HL data. 
From the values of the different parameters obtained through our best fit models (Fig.~\ref{eeuf_spectra_cenx3}), we measured 
a fraction of about 80\% and 30\% for the Comptonized cyclotron emission of the 
LL and the HL observation, respectively (see Fig. \ref{spectra_components} and Table \ref{tab_newcompmag}).

To perform a fit to the HL and LL data with the pencil-beam geometry (assuming a free-fall velocity profile 
for the material within the accretion column), we have to introduce the flow terminal velocity  
$\beta_0$ at the NS surface as an additional model parameter. 
We found that the fits to the data from both the HL and LL observations were unable 
to provide reasonable constrains to all model parameters. The situation did not improve by fixing the \compmag\ normalization, 
as different values of $\ncomp$ resulted in virtually equivalent \chiq. By performing fits with values of 
$\ncomp$ comprised between few tens and  a few thousands, we found that a tight relation exists between this parameter, 
the accretion rate $\mdot$, and the radius of the column $R_0$, i.e. $\mdot = a_1\ncomp^{-b_1}$ and  
$R_0=a_2 \ncomp^{-b_2}$ (as expected according to the parameter definitions; see Sect.~\ref{sec:new_compmag}). 
We measured the slopes $b_1=1.12\pm 0.03$ and $b_2=0.44\pm 0.01$ from the fits to the LL observation, $b_1=0.94\pm 0.02$ and $b_2=0.49\pm 0.04$ from 
the HL one.
By assuming for the pencil-beam geometry the same value of  
$R_0$ measured from the fan-beam geometry, we can use the above relations and coefficients to estimate the mass accretion rate 
in the former case. We obtained $\mdot \sim 1.6$ and $\mdot \sim 1.9$ for the LL and HL observation, respectively.
In the pencil-beam geometry, the spectral parameters related to the cyclotron feature and the continuum 
are consistent (to within the uncertainties) between the LL and HL observations. 
The electron temperature is slightly higher for both observations than that measured from the fits with the 
fan-beam geometry, but in both geometries a consistent decrease of $\kte$ from the LL to the HL observation  
is measured. From the fits, we found $\beta_0 <  0.1 $ and $\zmax \la 4$ km for either the LL and HL data.

\subsection{\fu}
\label{sec:fu}

In the case of \fu,\ we first tried to fit the spectra extracted from  Obs. I and II (see Table~\ref{tab:observations}) 
using an absorbed \compmag\ model in the fan-beam geometry, plus two absorption Gaussian lines at energies 
corresponding to the fundamental and the second harmonic of the CRSF. This gave \chiq/dof=710/422 and \chiq/dof=753/526 
for Obs. I and II, respectively.
In both cases, we noticed a systematic excess  at energies $\lesssim2~$keV
together  with an absorption feature around $\sim$ 35 keV 
(the latter were more significant in Obs. I). We thus first added a partial covering component to the 
previous spectral model (\pcfabs\ in \XSPECB), leading to  
\chiq/dof=518/420 and \chiq/dof=598/524 in Obs. I and II, and then added  to the model a further  Gaussian gaussian line to take into account the presence of the third CRSF harmonic; we obtained \chiq/dof=444/418 for Obs. I and \chiq/dof=583/522 for Obs. II, respectuvely.  
In both cases, the Gaussian widths of the third CRSF could not be constrained during the fit and thus we fixed this parameter   
to $\sigma=2$ keV, in agreement with the measured width of the second CRSF harmonic.  
In Obs. I, we additionally noticed some residuals from the fit around $\sim$45~keV, where the fourth CRSF harmonic 
is usually detected (see, e.g. F09). Adding a fourth Gaussian line to the spectral model used for Obs. I led to 
a further improvement in the fit with \chiq/dof=363/415. 

No evidence of the K$\alpha$ iron emission line was found in the spectra of Obs. I and II. 
We report all the best determined parameters of the spectral fits described above in Table \ref{tab_newcompmag}. 
The unfolded spectra and the residuals from the fits are shown  
in Fig.~\ref{eeuf_spectra_4u}. We also show in Fig.~\ref{spectra_components} 
the separated contributions of the Comptonized cyclotron and \bs\ emission to the total X-ray 
spectra of \fu; the former component is about 60\% and 70\% in
Obs. I and II, respectively.

Concerning the applicability of the \compmag\ model in the pencil-beam geometry to \fu\ (assuming a free-fall velocity 
for the material inside the accretion column), the same considerations of \cen\ apply (see Section \ref{sect:cen}).
From the fit to the source spectrum in Obs. I, we determined $b_1=1.06\pm 0.02$ and $b_2=0.52\pm 0.01$ for the two 
relations $\mdot \propto \ncomp^{-a}$ and $R_0 \propto \ncomp^{-b}$, respectively. 
The values measured from the fit to the source spectrum of Obs. II are  instead $b_1=0.93\pm 0.02$ and  $b_2=0.48\pm 0.01$.
The corresponding values of the mass accretion rate for a column radius $R_0$ equal to the value obtained in the fan-beam geometry are $\sim 0.76$ and $\sim 0.52$. Interestingly, the accretion rate ratio results equal to the luminosity ration of the two observations (see Table \ref{tab_newcompmag}).
At odds with the case of \cen, the flow terminal
velocity and column height determined for the two \fu\ observations remain roughly constant,  with $\beta_0 \sim 0.2$ and $\zmax \sim 3$ km.

\subsection{\her}

A fit with an absorbed \compmag\ model, together with a cyclotron absorption line, to the 
combined \suzaku\ and \nustar\ observation of \her\ provided an unacceptable result 
(\chiq/dof=7387/3732) due to a strong emission emission
line around 6.4-6.7 keV. The addition of an emission line  
at these energies led instead to a formally acceptable fit (\chiq/dof=3909/3729; see Table \ref{tab_newcompmag}). 
It is worth noting that F13 modelled similar residuals by using 
two narrow lines with centroid energies of 6.4 and 6.5 keV,
respectively (the estimated line widths were $\sigmag \sim 0.1$~keV). 
One of the two lines was included mainly to smear-out the sharp drop of the adopted 
phenomenological cut-off power-law model, and thus the second line is likely to have a model-dependent 
rather than a physical origin. 

Even though the fit was formally acceptable, some residuals were still visible 
around $\sim$10~keV. These residuals were also noticed by F13, who argued that they might 
have been caused by some instrumental effect and excluded the energy interval 10-12~keV from 
the spectral analysis. The fact that the same residuals emerged by using a complete different spectral 
model, convinced us that they could have instead a physical origin. We found that the addition 
of an iron absorption edge at 9.1~keV correctly accounted for the residuals around 10~keV and 
slightly improved the fit (\chiq/dof=3769/3727). 
The estimated contributions of the cyclotron and \bs\ Comptonized
component to the total X-ray emission of the source in the 3-100 keV energy range were 
60\% and 40\%, respectively. We report all values of the best fit model to the 
data of \her\ in Table~\ref{tab_newcompmag}, and show the unfolded spectrum with the residuals from the 
fit in Fig.~\ref{eeuf_spectra_her}. The high statistical quality of the \her\ observations allowed  
to constrain reasonably well all parameters of the \compmag\ model, providing the most accurate test 
of the model available so far. 

Concerning the details of the CRSF in \her,\ it is noteworthy that the width of this feature 
as measured by F13 was found to be dependent on the specific continuum model adopted by the authors and 
varied between 6 and 8~keV. The width of the CRSF estimated through the usage of our \compmag\ model 
is compatible with the value measured by F13 when a high-energy cut-off (\highecut\ in \xspec) is used to 
characterize the data. The latter is also found by F13 to be the most statistically preferable spectral 
model. The centroid energy of the CRSF measured by F13 and in the current paper are only marginally 
different ($\ecyc = 37.7-39.1$ keV and $\ecyc = 37.2\pm0.2$~keV, respectively). 

Finally, in the case of \her\ it was not possible to 
find a formally acceptable fit to the data by using the \compmag\ model 
in the pencil-beam geometry and considering a free-fall velocity profile for the material 
within the accretion column. 

\begin{table*}
\caption{Best fit models for the sample of sources analysed in this paper. 
For the \compmag\ model, the emission geometry defined in equation (\ref{flux_geom2_b}) 
combined with velocity profile in equation (\ref{vel_2}) is assumed.}
\begin{center}
\scriptsize
 \begin{tabular}{ l c c c c c }
\hline
\hline
Source    & \cen (LL)  & \cen (HL)   & \fu\ (Obs. 1)  & \fu\ (Obs 2)  &  \her \\
Parameter    &    &     &    &   & \\
\hline
\multicolumn{6}{c}{ \sc{Continuum and derived parameters} } \\
\hline
\smallskip
\smallskip
$\Nh$\tablefootmark{a}    &   0.65$^{+  0.17}_{-  0.13}$   & 0.71$^{+  0.06}_{-  0.07}$    & 1.12$^{+  0.07}_{-  0.07}$  & 1.19$^{+  0.09}_{-  0.10}$    &  $[10^{-4}]$ \\
\smallskip
$\Nh$\tablefootmark{b}     &   6.3$^{+  0.9}_{-  0.8}$   &  --   & 5.3$^{+  1.0}_{-0.9}$   & 4.3$\pm0.9$   & --\\
\smallskip
Cov. Fract.    & 0.76$^{+  0.02}_{-  0.03}$ & --    &  0.54$^{+  0.03}_{-  0.03}$  & 0.51$^{+  0.04}_{-  0.05}$  & --\\
\smallskip
$\kte$ (keV)   & 1.83$^{+  0.04}_{-  0.04}$   &  1.43$^{+  0.03}_{-  0.03}$   &  0.77$^{+  0.05}_{-  0.03}$  & 0.89$^{+  0.04}_{-  0.02}$  &   3.09$^{+  0.10}_{-  0.12}$ \\
\smallskip
$\mdot$\tablefootmark{c}    & 0.83$^{+  0.07}_{-  0.06}$   &  11.7$^{+  2.0}_{-  1.4}$  & 0.92$^{+  0.13}_{-  0.20}$   & 0.47$^{+  0.05}_{-  0.10}$ &   4.02$^{+  0.40}_{-  0.30}$ \\
\smallskip
$B_{12}$\tablefootmark{d} &  2.62$\pm0.09$   &  2.51$\pm0.06$   & 0.99$\pm0.01$  &   0.97$\pm0.01$  & 3.22$\pm0.02$   \\
\smallskip
$\sigmacyc$ (keV) &  [3]  &  1.6$\pm0.4$  & 2.4$\pm0.3$  & 2.3$\pm0.2$ &  3.1$\pm0.4$  \\
\smallskip
$R_0$~(km) &  1.02$^{+  0.06}_{-  0.05}$  &  1.72$^{+  0.07}_{-  0.06}$   & 0.44$^{+  0.03}_{-  0.04}$  & 0.34$^{+  0.01}_{-  0.03}$ &  1.70$^{+  0.16}_{-  0.12}$  \\
\smallskip
$H$~(km) &  7.3$\pm0.4$  &   3.4$\pm 0.2$  & 2.5$\pm 0.1$  & 2.6$\pm 0.1$ &    4.6$\pm0.3$ \\
\smallskip
$N_{\rm comp}$ & [65]    &  65$\pm5$   &   4.4$^{+1.1}_{-0.7}\times10^3$ &   4.4$^{+2.0}_{-0.7}\times10^3$   &  13$\pm2$ \\
\hline
\multicolumn{6}{c}{ \sc{Cyclotron absorption features}} \\
\hline
\smallskip
$E_1$~(keV) & 30$\pm1$   &  29.1$\pm0.7$   &  12.2$\pm 0.1$  & 11.9$\pm 0.1$ &  37.2$\pm0.2$ \\
\smallskip
$\sigma_1$~(keV) & [5]   &   6.0$^{+  1.0}_{-  0.8}$  & 1.7$\pm 0.2$   & 1.5$^{+  0.2}_{-  0.1}$ & 5.7$^{+  0.6}_{-  0.2}$ \\
\smallskip
$N_{1}$\tablefootmark{e} & 9$\pm2$   &  7.4$^{+  2.5}_{-  1.7}$   &   0.9$^{+  0.2}_{-  0.1}$ & 0.6$\pm 0.1$ &   9.5$^{+  3.3}_{-  1.4}$  \\
\smallskip
$E_2$~(keV) & --   &  --   &  23.0$\pm0.2$  & 22.6$\pm0.1$ & -- \\
\smallskip
$\sigma_2$~(keV) & --   &  --   &  2.1$^{+  0.4}_{-  0.3}$  &  2.0$\pm 0.2$   & --  \\
\smallskip
$N_{2}$\tablefootmark{e} & --   &   --  &  1.3$^{+  0.3}_{-  0.2}$  & 1.3$^{+  0.1}_{-  0.2}$ &  -- \\
\smallskip
$E_3$~(keV) &  --  & --    &  35.2$\pm0.6$  & 36.0$^{+  1.7}_{-  1.3}$ &  --  \\
\smallskip
$\sigma_3$~(keV) & --   & --    &  [2]  & [2] &    --     \\
\smallskip
\smallskip
$N_{3}$\tablefootmark{e} & --   &  --   & 1.3$^{+  0.2}_{-  0.3}$  & 0.4$\pm 0.2$ &    --      \\
\smallskip
$E_4$~(keV) &  --  & --    &  46.4$^{+  1.0}_{-  0.9}$ & -- &  --  \\
\smallskip
$\sigma_4$~(keV) & --   & --    &  3.6$^{+  1.7}_{-  1.4}$   & -- &    --     \\
\smallskip
\smallskip
$N_{4}$\tablefootmark{e} & --   &  --   &  2.0$^{+  0.7}_{-  0.5}$   & -- &    --      \\
\hline
\multicolumn{6}{c}{ \sc{Iron emission line}}    \\
\hline
\vspace{0.2cm}
$E_{\rm Fe}$~ (keV) & 6.58$\pm0.01$   &  6.60$\pm0.03$   &  --  & --  &  6.61$\pm0.03$  \\
\smallskip
$\sigma_{\rm Fe}$~ (keV)& 0.33$\pm0.02$   & 0.31$\pm0.05$    & --   & -- &  0.35$^{+  0.10}_{-  0.05}$    \\
\smallskip
$N_{\rm Fe}$\tablefootmark{f} & 5.2$\pm0.2$ &   8.7$^{+  1.1}_{-  0.9}$  & --   & -- &   5.7$^{+  1.7}_{-  1.0}$ \\
\hline
\smallskip
$F_{\rm cyc}/F_{\rm tot}$\tablefootmark{g} &  0.78  &  0.28   & 0.58   & 0.67 &  0.62     \\
\smallskip
\smallskip
$L_{\rm X} $\tablefootmark{g,h} &  1.2  &  11.0   & 11.7   & 7.8 &    4.6     \\
 \chiq/dof & 449/312   &  167/238   & 363/415   & 583/522  &  3769/3727       \\
\hline
\hline
\end{tabular}
\end{center}
\tablefoot{
\tablefoottext{a}{Absorption of the \phabs\ model in units of $10^{22}$ cm$^{-2}$.}
\tablefoottext{b}{Absorption of the \pcfabs\ model in units of $10^{22}$ cm$^{-2}$.}
\tablefoottext{c}{In units of $1.5\times 10^{17}$ $m$ gr~s$^{-1}$, with $m=1.4$.}
\tablefoottext{d}{In units of $10^{12}$ G, and linked to the centroid of the scattering features: the fundamental for \her and \cen, the first harmonic for \fu.}
\tablefoottext{e}{Normalization corresponding to the parameter $\tau_{\rm n}$ in equation (\ref{eq:gaus}).}
\tablefoottext{f}{In units of $10^{-3}$ photons cm$^{-2}$.}
\tablefoottext{g}{Computed in the  energy range 0.5-100 keV for \cen\ and \fu,  3-100 keV for \her, and corrected for absorption.}
\tablefoottext{h}{In units of $10^{37}$ erg s$^{-1}$ assuming a distances of 8 kpc, 7.5 kpc, and 6.5 kpc for \cen, \fu, and \her, respectively.}
}
\label{tab_newcompmag}
\end{table*}

\section{Discussion}
\label{sec:discussion}

In this paper, we introduced an updated version of the \compmag\ model, which permits 
to physically compute the X-ray luminosity emerging from a highly magnetized accreting pulsar in 
the fan-beam and pencil-beam geometry. 

Two particularly critical parameters of the model are the normalization $\ncomp$ and the 
mass accretion rate $\mdot$. In a purely Newtonian case, the mass accretion rate entirely 
regulates the source X-ray luminosity and it is expected that the normalization of the \compmag\ 
model can be fixed to a constant proportional to the source distance (see Sect.~\ref{sec:results}). 
In a more realistic case, the X-ray emission produced relatively close to the NS surface will be strongly 
affected by General Relativistic (GR) effects and the normalization of the \compmag\ model is no longer trivially 
connected to the source distance. As in this work the \compmag\ model was tested against observational 
data in which GR effects cannot be disentangled from the local X-ray emission, we left $\ncomp$  
free to vary in the fits and determined its value for the different sources from the data. A better definition of 
$\ncomp$ would require the extension of the \compmag\ model to include a full GR treatment of the X-ray 
emission from the pulsar, but this is outside the scope of the present paper. 

The simplified approach adopted here could be successfully applied to the currently best available 
broad-band X-ray observations of three bright X-ray pulsars: \fu,\ \cen,\ and \her.\ 
We used publicly available \sax\ archival data for \cen\ and \fu,\ and a recent combined  
\suzaku+\nustar observation of \her.\ 
In all cases, the broad-band coverage of the data ensured a detailed characterization of
the source spectrum and provided the possibility to perform tests of the \compmag\ model.
In the case of \fu, the fit performed to the two BeppoSAX observations with $\ncomp$ and $\mdot$ left 
free to vary provided fully self-consistent physical results. We obtained a ratio of 
$\sim 1.3$ between the mass accretion rate in Obs. I and Obs II, and a ratio of $\sim 1.8$ between the corresponding 
values of $\ncomp$. For comparison, the ratio between the X-ray luminosities recorded during these observations 
is $\sim 1.5$ (0.1-100~keV energy range; see Table~\ref{tab_newcompmag}). 
On one hand, the difference in the mass accretion rate evaluated though \compmag\ thus fully reflect the variation 
of the source luminosity across the two data-sets. On the other hand, the difference in the pulse profile of the source 
that was reported previously among these two observations \citep[][F09]{santangelo1999} also justify the measured 
variation in $\ncomp$, as some change has certainly occurred in the pattern of the X-ray radiation arising from the pulsar 
accretion column. A similar conclusion applies to the case of \cen,\ for which a ratio of $\sim 7.3$ was measured from the fits 
with \compmag\ between the mass accretion rats in the LL and HL observations and the correspondingly  known ratio of the X-ray luminosity is 
$\sim 8$ (see Table~\ref{tab_newcompmag}). Note that in this case the value of $\ncomp$ of the LL observation had to be 
fixed to that of the HL observation in order to avoid degeneracies between this parameter and $\mdot$ in the spectral fits 
(see Section \ref{sect:cen}). A similar check was not possible on \her,\ as in this case only a single observation 
was available. However, the \compmag\ model could provide also for this source a reasonably good fit to the broad-band 
X-ray data, giving indications about the structure of the accretion column (see Table~\ref{tab_newcompmag}).  
Generally, we remark that a comparison between the values of $\mdot$ determined from \compmag\ for different systems  
should be carried out with caution, as uncertainties on the distances to these sources and intrinsic differences
in the accretion geometry could lead to systematic biases. Values of $\mdot$ are more easily comparable between 
different observations of a single source, as in this case the uncertainty on the distance cancels out and variations 
in this parameters can be directly linked to the physical processes intervening in the release of the X-ray emission. 

There are two major limitations in the current version of the \compmag\ model. 
The first is related to the crude approximation of the scattering cross-section, which
is described by parallel and perpendicular components to the magnetic field direction 
($\sigma_\parallel$ and $\sigma_\perp$, respectively). The former appears in the Fokker-Planck treatment 
of the radiative transfer equation (\ref{fp_eq}) 
through the vertical spatial photon diffusion term, while the
latter is related to the photon escape time across the side walls of the
accretion column. The cross-section $\bar\sigma$, related to the photon
diffusion in the energy space and representing the average scattering
direction of photons parallel and perpendicular to the magnetic field, is also crudely approximated. 
These issues were already pointed out by BW07\footnote{Note that BW07 treats the parallel and average cross-sections 
as implicit model parameters, while in the \compmag\ model we fix them to a suitable values.} 
and described in the first version of the \compmag\ model 
by \citet{farinelli2012a}. We stress them here once more for completeness. 

As also briefly mentioned above, the second important limitation of the \compmag\ model is 
the usage of a full Newtonian treatment to compute the pulsar X-ray emission. As the latter is emerging 
close to the NS surface, it is likely that GR will play a key role in shaping the final energy distribution 
and intensity of this emission before it gets to the observer. The different geometrical parameters 
inferred from the \compmag\ model, as the accretion column radius $R_0$ and its height $H$, would probably need to be 
scaled in a non-trivial way to their values in the NS reference frame. This is, however, a limitation that also applies 
to several spectral models available within \XSPECB\ and commonly used in the X-ray astrophysical community. 
A well know case is that of the \diskbb\ model \citep{mitsuda84}, which assumes a Newtonian geometry and is widely 
used to characterize the soft spectral component of X-ray binaries hosting NSs and black-holes. 
It is noteworthy that the accretion column radius derived from the application of the \compmag\ model to the cases of 
\fu,\ \her,\ and \cen,\ provides a value of $R_0 \sim 1$ km that is qualitatively in agreement with that expected 
in case the accreting material is led by a relatively intense magnetic field before arriving close to the NS surface 
\citep[$B\sim10^{12}$\,G; see, e.g.,][]{basko1976}.  

Finally, we discuss the usage of the RTE (equations~[\ref{fp_eq}] and [\ref{fp_eq_b}]).  
In the present case, the RTE is written in the static reference frame of the observer with
an accuracy of order $\beta$, the adimensional bulk velocity of the accreting material.
It is important to outline that for a decelerated velocity profile
as in equation~(\ref{vel_2}), the electron density progressively increases
towards the NS surface. 
As the combined continuum \bs\ plus narrow cyclotron emission is proportional
to $\ne^2$ (see equations~[\ref{cyc_emission}] and [\ref{bs_emission}]), most
of the seed photons are produced in the region where $\beta \rightarrow 0$.  
Corrections between the reference frames of the fluid and the observer are thus  minimal.
The same consideration holds for the escape time of photons and
the Comptonization efficiency: both processes are more efficient
in a high electron-density environment. 
This also significantly reduces the uncertainty in equation~(\ref{fp_eq}) introduced by the 
usage of the relation $K/J=1/3$ between the second and zeroth moment of the intensity
field. The latter would strictly hold only in the fluid reference frame (see Appendix~\ref{appendix_b}).

\subsection{Comparison with earlier models}

As already reported in Section \ref{sec:new_compmag} and with more details 
in F12, there are a number of differences between the current implementation
of the \compmag\ model and the former analytical treatment presented
by BW07. So far, the only application\footnote{An application of the BW07 model to other sources was preliminary  
discussed in a conference proceedings by \citet{marcu2015}.} of the BW07 model to a broad-band spectrum 
of an X-ray pulsar was presented by F09 using BeppoSAX data of \fu.\ 
We thus compare below the results of the present paper for the source \fu\ with those reported 
in F09. 

A first important difference between the two treatments is that the BW07 model was proved unable to provide 
a satisfactorily fit over the entire energy range spanned by the X-ray spectrum of \fu.\
F09 showed that an additional thermal Comptonization component was needed 
below $\sim$9~keV to correctly describe the data, while the higher energy part could be 
well interpreted in terms of Comptonized cyclotron emission. A broad Gaussian line had also to be 
included in the spectral model to achieve an acceptable result, but the origin of this feature could not 
be easily explained. 
F09 also showed that the magnetic field strength derived from the centroid energy of the cyclotron line was 
significantly different from that inferred from the fit to the continuum, and the values of $\mdot$ and 
$R_0$ had to be fixed in the fit ($6\times10^{16}$~g~s$^{-1}$ and 600~m, respectively) 
due to unresolved degeneracies of the different model parameters. 
As we showed in the previous sections, the \compmag\ model is able to consistently fit the broad-band 
spectra of the different sources without the need of including additional components 
(besides the cyclotron absorption features). 

It should be remarked that the lower electron temperature of the \bs\ seed component measured in the \compmag\ and BW07 models 
($\kte \sim 8$\,keV in F09 and $\kte \sim 0.7$\,keV in the present work) is known to arise 
from the different approximations of the Compton scattering: BW07 considered only the first-order 
bulk Comptonization term in the energy-diffusion operator, while we introduced also the second-order term.
The latter component, being space-dependent, can not be included when
performing the analytical treatment of Comptonization through the
variable-separation method (as adopted by BW07). It is a general feature of the thermal plus bulk 
Comptonization models that the inclusion of higher-order Compton bulk terms lead to lower electron
temperatures for the same spectrum \citep{titarchuk1997,psaltis97}.

In our analysis of the X-ray emission from \fu,\ we did not find any statistically significant evidence 
of the fifth harmonic of the cyclotron line, that was instead reported by F09. 
We note, however, that the parameters of the cyclotron lines are known to be strongly affected by the choice 
of the X-ray continuum and thus the detection of particularly weak wiggles in the high-energy end of 
an exponentially decaying spectrum can easily turn out to be statistically non-significant in different models.

\section{Source term emission}
\label{sec:source_terms}

In this section we justify the assumption of a combined thermal \bs\ and cyclotron 
emission (the latter approximated through a Gaussian function) to generate the seed photons 
used for the Comptonization in the \compmag\ model. 
We mentioned in Sect.~\ref{sec:intro} that this approximation was originally proposed by BW07, 
given the current impossibility of providing a self-consistent analytical description of the 
magnetic \bs\ emission produced by the electrons in the accretion flow. While implementing the BW07 
treatment of the problem in an \XSPECB\ model, F09 adopted for the cyclotron term of the seed photons  
an exponentially attenuated $\delta$-function. These authors thus assumed that the width of the 
cyclotron line is intrinsically narrow at the origin and is then broadened due to thermal and bulk 
Comptonization effects. In the treatment presented in this paper, we used a Gaussian profile 
instead of a $\delta$-function in order to take into account the natural width of the  
cyclotron emission feature (the \compmag\ model parameter describing this effect is 
$\sigma_{\rm cyc}$ in equation~[\ref{source_cyc}]). The latter is indeed expected to result from both 
thermal broadening and magnetic field mixing instabilities. 

\subsection{Bremsstrahlung emission}
\label{sec:brem}

The fits to the data of the X-ray pulsars considered in this paper with the \compmag\ model 
provided an electron temperature of order of few keV (see Tab.~\ref{tab_newcompmag}).  
It is straightforward to verify for each source 
that the fraction of electron  with velocities $|v| > \sqrt{2 E_{\rm cyc}/\me}$
is very small (less than 1\%)  in both the cases of 1D and 3D Maxwellian 
distributions. The effect of tail depopulation because of the resonant electron-proton
interaction with photon emission, and the consequent deviation of
$f(v)$ from a Maxwellian distribution is thus very marginal.
Moreover, as previously mentioned in Section \ref{sec:new_compmag}, the angular integration of the \bs\ emission  
smears-out narrow features arising from the effect of the viewing angle, 
resulting in a spectrum which fairly well agrees with the three-dimensional classical case. 
We thus concluded that splitting the \bs\ emission process in a strong magnetic
field into a smooth continuum similar to the unmagnetized case plus a
narrow cyclotron feature is a reasonably good assumption.

It shall be noticed that, beside the proton-electron \bs,\ there is a second channel for the production of 
cyclotron emission photons due to the photon-electron resonant scattering \citep[RS; see, e.g.,][]{canuto71, herold79, 
daugherty86, harding91}. In principle, for a thermal plasma with a temperature of a few keV, only a small fraction of 
the electrons fulfill the condition $E_{\rm el} > E_{\rm cyc}$. The presence of a significant cyclotron feature 
in the X-ray data (see Fig.~\ref{spectra_components}) would thus support the idea that the e-$\gamma$ RS dominates 
over the e-p RS. In reality, a quantitative estimate of the relative contribution provided by the two
processes is very difficult. The \bs\ spectra reported by  R99
were computed in the case of an optically thin plasma, where 
the e-p processes dominate over the e-$\gamma$ interactions.
A more realistic solution to the problem would require a
self-consistent treatment where both processes are taken simultaneously into account,
employing a non-linear iterative solution scheme. 
The possibility of exchanging the \bs\ and cyclotron terms for the seed photons in equation~\ref{fp_eq}
indicates our inability in developing such scheme. 

\subsection{Cyclotron emission}
\label{sec:cyc}

Let us now consider the broadening of the cyclotron emission feature due to magneto-hydrodynamical effects.
First, we emphasize that the cyclotron emission term in the \compmag\ model 
(equation~[\ref{source_cyc}]) is  self-consistently weighted over the 
vertical profile of the accretion column, as we considered $B_{z}=B_0 (z_0/z)^3$. 
An intrinsic broadening of the cyclotron emission line energy $E_{\rm cyc}$ is thus 
possible due to geometrical effects (i.e., the distance from the NS) 
and the assumption on the vertical emissivity profile of the accretion column. 
The vertical gradient of the magnetic field does not provide, however, a dominant contribution 
to the broadening of the cyclotron emission feature, as the total height of the column 
where the overall emission process takes place is relatively small 
(see Tab.~\ref{tab_newcompmag}) and cannot lead to $\sigma_{\rm cyc}$ appreciably
larger than zero. 
A similar conclusion holds for the radial gradient of the magnetic field. 
The exact expression of a dipolar magnetic field in the vertical direction using cartesian coordinates is  
\begin{equation}
 B_{\rm z}=\mu \left[ \frac{2z^2}{(R^2 + z^2)^{5/2}}- \frac{1-\frac{z^2}{R^2+z^2}}{(R^2 + z^2)^{3/2} }\right],
\end{equation}
where $\mu$ is the magnetic moment, and $R^2=x^2+y^2$.
The assumption $B_{z} \propto z^{-3}$ considered so far is thus valid only in the asymptotic 
limit $R \ll z$. However, it is easy to demonstrate that the variations of both $B_{z}$ and $\vert B \vert$ from the
center ($R=0$) to the walls of the accretion column ($R \la 1$~km) are negligible and cannot contribute 
significantly to $\sigma_{\rm cyc} > 0$.

Magneto-hydrodynamical simulations of hyper-critical accretion regimes
onto a magnetized star have shown that significant fluctuations of the magnetic 
field intensity in the vertical direction can be produced as a consequence of field 
submerging effects by the copious inflowing material in the accretion column \citep{bernal2010}.
Numerical simulations of plasma accreted on the NS surface also show that
interchange instabilities contribute to increase the gradient of the magnetic field 
intensity especially at the base of the accretion column \citep{mukherjee2013a,mukherjee2013b}. 
We thus argue that all these effects can easily give rise to the complex magnetic field structures 
that are needed to produce the observed broad cyclotron emission feature close to the NS surface. 

It is also interesting to discuss the relative contribution of the Comptonized cyclotron
and \bs\ components to the total emission of the analyzed sources, even though the two terms 
for the production of the seed photons for the Comptonization should be mutually linked in a more realistic 
treatment (see Sect.~\ref{sec:new_compmag} and the discussion earlier in this section). 
As shown by \cite{neugebauer96} through detailed calculations, if the energy of an electron
moving along a magnetic field line is higher than the cyclotron energy, the 
proton-electron Coulomb cross-section has four resonances.
The first two occur at $v'= \pm v$, where $v$ and $v'$ are the 
velocities of the electron before and after the scattering, respectively.
In these cases the p-e scattering process does not lead to the emission of a photon 
and the forward scattering probability is much higher than the backward one due to the 
high proton mass (making this a nearly pure elastic scattering). 
The other two resonances occur at $\pm v'$, such that $E'=(E_{\rm el}- \ecyc)$. 
Here $E=1/2 \me v^2$, $E'=1/2 \me v'^2$, and the $\pm$ sign refers 
to the forward and backward scattering of the electron, respectively.
In this case, the parallel velocity component (and thus the energy) is transferred to the
quantized perpendicular electron Landau level, from which the electron decays
releasing a cyclotron photon within a characteristic time
\begin{equation}
t_{\rm cyc} \sim 10^{-19} \left(\frac{B_{\rm crit}}{B}\right)^2 {\rm s}, 
\end{equation}
\noindent
where $B_{\rm crit}=4.41 \times 10^{13}$ G.  
The p-e scattering leading to the emission of a photon is the most interesting 
process from an observational point of view. 
In particular, the resonance occurs because the electron in the
intermediate state (before the decay) is generally in the excited Landau levels with $n=1$. 
A p-e scattering leading to a photon emission and an electron in the excited Landau level with $n=0$
during the intermediate state is also a possibility, but the photons released after the decay 
would have in this case a lower energy than the cyclotron value. This case is usually 
considered similar to a zero-magnetic-field \bs\ emission.
From the represented Comptonized cyclotron and \bs\ emissions in Fig.~\ref{spectra_components},  
we conclude that for all sources considered in this work the bulk of the \bs\ emission
is produced at energies lower than the cyclotron emission, in line with the
above discussion. The contributions of the two components to the total 
emission of each source are similar, as expected in case of p-e Coulomb scatterings where the electrons 
are populating (for the cyclotron component) or not (for the \bs\ component) the Landau levels $n \ge 1$ 
during their intermediate excited state. 

\begin{figure}
\includegraphics[width=6.5cm, angle=270]{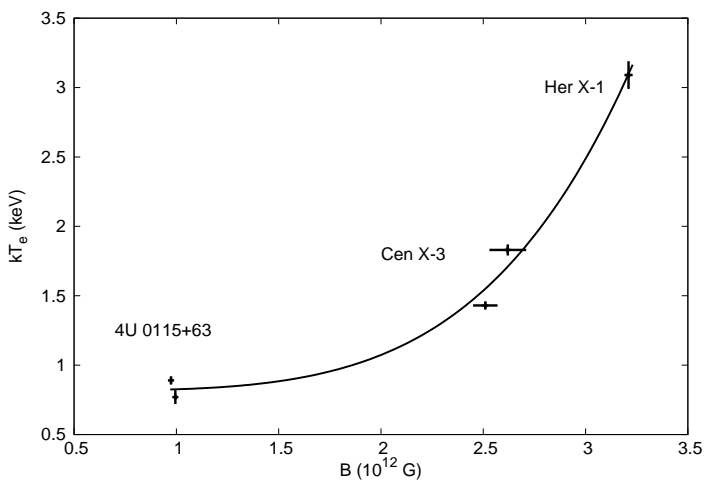}
\caption{Electron temperature as a function of the magnetic field at the base of the column 
for the sample of accreting X-ray pulsars analyzed in this paper. The data values are reported in 
Table~\ref{tab_newcompmag} and can be fit by a powerlaw (see text for more details).}
\label{fig:kte_vs_b}
\end{figure}

It is also worth noting the strong correlation between the electron temperature and the magnetic field 
intensity derived from the fits with the \compmag\ model to the data of all sources  (Fig.~\ref{fig:kte_vs_b}). 
The Spearman  correlation coefficient of the dataset is $R=0.9$, and  the relation between the two parameters 
can be analytically described by the function $\kte=a  + b \b12^c$, with $a=0.81 \pm 0.10$, $b=0.01 \pm 0.008$, and $c=4.61 \pm 0.70$.  
As the electron temperature in the \compmag\ model determines the exponential roll-over of the spectra, the correlation 
we found is most likely related to the scaling law between the cut-off energy and the centroid energy of the cyclotron emission 
lines that was discovered thanks to Ginga and RXTE data through fits with phenomenological 
models \citep[see, e.g.,][]{Makishima1999,coburn2002}. The advantage of the \compmag\ model in this case is that it can 
provide a physical interpretation of this correlation. 
In particular, we note that the electron temperature is about an order of magnitude lower than the cyclotron energy,
which provides the maximum temperature at which electrons can be heated. As discussed earlier in this section, this is due 
to the collisional excitation of the Landaus levels, which limits the electron Maxwellian distribution below the cyclotron energy.
As reported by \citet{Arons1987}, the cyclotron emission is an effective cooling channel for the material in the 
accretion column (together with the Compton cooling). This is also the reason why there is an important contribution 
of the cyclotron emission term in the spectra of all X-ray pulsars analyzed in this paper. \citet{Arons1987} also
showed that the electron temperature is expected to be generally lower than the average photon energy,
because photons have the largest heat capacity. This can be easily verified for \fu,\ \her,\ and \cen\ by using the 
results reported in Table~\ref{tab_newcompmag}. 

While the relation between the magnetic field intensity and the electron temperature is thus well understood 
in terms of the physical processes regulating the accretion in highly magnetized X-ray pulsars, the 
possible correlations we identified in Sect.~\ref{sec:results} between the different geometrical parameters of the 
\compmag\ model (i.e., $R_0$ and $\zmax$) should instead be taken with caution. These relations were derived 
by assuming a purely Newtonian treatment of the X-ray emission process close to the NS surface, where GR effects 
(e.g., light bending and lensing) can significantly modify the values of geometrical quantities between 
the emission and the observer frame. In order to be properly understood and interpreted, the same correlations 
should be investigated by introducing a ray-tracing code providing solutions for the equations describing the 
geodesics of the light emerging from the accretion column. This is, however, outside the scope of the present paper. 

\section{Geometry of the emission region}
\label{sec:geom}
 
In modelling the source spectra, we have considered both a pencil-beam
geometry with an accelerating velocity profile in the accretion column, 
and a fan-beam geometry, where matter decelerates towards the NS surface.
The first case is expected to occur 
in less luminous sources ($L_{\rm X}  < 10^{37}$ erg s$^{-1}$), 
where the radiation pressure generated by the soft photons from the NS surface 
is too low to halt the infalling matter in the accretion column 
and the radiation is preferentially emitted from the top of the column (fan-beam). 
At higher luminosities ($L_{\rm X}  > 10^{37}$ erg s$^{-1}$), 
the strong radiation pressure from the NS surface contrasts 
the pressure of bulk inflow, leading to the formation of a radiative shock above the NS surface. 
In this case, the energy dependency of the electron-photon scattering cross section 
leads to substantially different behaviours below and above the 
cyclotron line energy. 
Below the cyclotron energy, the optical depth perpendicular to the magnetic field
is higher than in the parallel direction. The Comptonized radiation thus emerges preferentially 
through the lateral walls of the accretion column (fan-beam), while part of the weakly reprocessed 
radiation can still be emitted along the column.  
Above the cyclotron energy, the cross sections in both directions become comparable, increasing the 
emission through the lateral walls of the column. The \compmag\ model is 
best suited to fit phase-averaged spectra, as in this case the effects described above 
are averaged out and the treatment of the cross sections in the RTE approximation is justified. 

All the sources considered in this paper have X-ray luminosities $L_{\rm X}  > 10^{37}$ erg s$^{-1}$, 
and thus the usage of the fan-beam geometry in the \compmag\ model was considered more appropriate 
to fit their X-ray spectra. We have shown that in principle fits with the \compmag\ model and the 
pencil-beam geometry are possible for \cen\ and \fu, while no solution could be found in the case of 
\her. In the pencil-beam case, the empirical dependence $R_0 \propto \ncomp^{-1/2}$ discussed 
in Sect.~\ref{sect:cen} and \ref{sec:fu} is derived from the definition of the 
\compmag\ normalization for the top-column emission (see equation~[\ref{flux_geom1_adim}]).
Indeed, the $\ncomp$ parameter explicitly depends on the emitting area at the top of the accretion column 
as well as on the magnification factor $\fgeom$. The latter parametrizes all the (unknown) geometrical 
and GR effects that are not taken into account in the present version of the \compmag\ model. 
The inverse proportionality $\mdot \propto \ncomp^{-1}$ shows that the \compmag\ model correctly 
takes into account a linear dependence between the source flux and the estimated mass accretion rate.
For a given value of the source X-ray flux, a decrease of $\ncomp$ has to be compensated 
by a corresponding linear increase in $\mdot$. We verified that the changes in $\mdot$ and $R_0$ 
derived as a consequence of a variation in the model normalization, $\ncomp$, are sufficient 
to keep  value of the \chiq unchanged. As we reported during the analysis of \cen\ and \fu,\ both
$\beta_0$ and $\zmax$ are insensitive to variations of $\ncomp$. This can be explained considering that 
the emerging flux (and thus the source X-ray luminosity) in the Eddington approximation (see equation~[\ref{flux_geom1}]) 
is mostly regulated by the vertical space gradient of the photon field at the boundary of the 
accretion column and not by the height of the column $\zmax$.
Although the value of $\partial n/\partial\tau$ at the top of the accretion column
is tightly related to the overall space photon distribution, the effect 
of the value of $\zmax$ is marginal in the pencil-beam case compared to the 
fan-beam case. In the latter, this parameter explicitly appears in the integration
of the source flux performed along the column height (see equation~[\ref{flux_geom2_b}]).
In this case, $\zmax$ influences more the vertical gradient of the magnetic 
field and thus also the cyclotron emission term. 

On one hand, it is thus evident that we should generally expect some degeneracy in the 
compmag model parameters and it might not always be possible to verify {\it a priori} 
if the fan-beam or the pencil-beam geometry is better suited to describe the X-ray spectrum 
of an X-ray pulsar. On the other hand, the fact that the pencil-beam geometry could not be used to 
satisfactorily fit the X-ray spectrum of \her\ gives us confidence that the \compmag\ model 
might be able to distinguish between the different emission states of these sources, at least in 
case high statistical quality and wide broad-band data are available. 
Additional observations of these sources with the next generation of X-ray instruments providing 
broad-band X-ray coverage \citep[as those on-board Astro-H; see, e.g.,][]{astroh} 
will give us the possibility of performing additional tests of the 
\compmag\ model and improve the treatment of all physical processes currently included in the 
corresponding code.

\section{Conclusions}
\label{sec:conclusions}

In this work we presented an updated version of the \compmag\ model 
and applied it to the case of three bright X-ray pulsars for which broad-band high statistical 
quality data were available. 
We found that the model is well suited to describe quantitatively 
the X-ray spectral energy distribution of these systems,
providing reasonable values of the physical and
geometrical parameters that are used to compute the high energy emission of the different sources. 
We widely discussed the limitations and the difficulties of the computations performed 
within the model. These are mainly related to the usage of vertically
and horizontally averaged photon-electron scattering cross sections, together 
with a cylindrical approximation treatment in which all
quantities vary only along the vertical coordinate.
For these reasons, the model can be better exploited to carry out the analysis 
of phase-averaged spectra. 
Despite its limitations, the \compmag\ model represents an additional significant step 
forward in the study of highly magnetized accreting X-ray pulsars. 

\begin{appendix}

\section{Derivation of the \bs\ and cyclotron source terms in the RTE}
\label{appendix_a}

The  cyclotron and \bs\ emissivity in units of erg cm$^{-3}$ s$^{-1}$ Hz$^{-1}$
are
\begin{eqnarray}
\label{cyc_emission}
\Psi_{\rm cyc}(E)=3.9 \times 10^{-38} \ne^2 B^{-3/2}_{12} H\left(\frac{E_{\rm cyc}}{\kte}\right) \\\nonumber
 \times e^{-\ecyc/\kte}  \delta(E-\ecyc),
\end{eqnarray}
where 
\begin{displaymath}
\label{h_funct}
H\left(\frac{E_{\rm cyc}}{\kte}\right)\equiv \left\{ \begin{array}{ll}
0.41 & \ecyc/\kte \geq 7.5\\
0.15~\sqrt{\ecyc/\kte} & \ecyc/\kte < 7.5. 
\end{array}\right.  
\end{displaymath}
and 
\begin{equation}
\label{bs_emission}
\Psi_{\rm bs}(E)=6.8 \times 10^{-38} \rho^2 T^{-1/2}_{\rm e}  e^{-E/\kte},
\end{equation}
with the temperature $T_{\rm e}$ expressed in Kelvin.

The units of the radiative transfer equation (\ref{fp_eq}) are s$^{-1}$, while the occupation 
number is related to the specific intensity by
\begin{equation}
\label{occ_numb_a}
J(E)=\frac{2 h \nu^3}{c^2} n(E),  
\end{equation}
where $J(E)$ is given in units of energy cm$^{-2}$ s$^{-1}$ Hz$^{-1}$ ster$^{-1}$.
Note that in equation~(\ref{occ_numb_a}) both $J(E)$ and $n(E)$ are averaged over the 
angle. 
To obtain the correct units for the equation (\ref{fp_eq}), the quantity
$J(E)$ must be multiplied by the extra factor $c$.

Using the definitions $E=h \nu$ and $x=E/\kte$, together with the 
equation (\ref{occ_numb_a}), the cyclotron and \bs\ source terms 
in the forms used within the equation (\ref{fp_eq}) can be easily 
obtained as 
\begin{equation}
\label{occ_numb_b}
j(x)= \frac{c^3 h^2}{2 (x \kte)^3} \Psi(E).
\end{equation}

Note however that the \compmag\ model solves equation (\ref{fp_eq_b}) which
is obtained  dividing equation (\ref{fp_eq}) by $\ne \sigma_\parallel c H$.
By additionally defining $\Theta=\kte/\me c^2$, we finally obtain the expression
\begin{equation}
\label{occ_numb_c}
\mathcal{S}(x)= \frac{c^2 h^2 }{2 (\me c^2)^3 x^3 \Theta^4 \ne \overline{\sigma}} \Psi(E),
\end{equation}
which provides the source term defined
in equations (\ref{source_cyc}) and (\ref{source_bs}), once the proper values of the constants are used.
Note also that in order to derive the expression for the cyclotron
term, we first used the property of the Dirac $\delta$-function 
\begin{equation}
\delta(E-\ecyc)=\frac{\delta(x-x_{\rm cyc})}{\kte},
\end{equation}
and then substituted  $\delta(x-x_{\rm cyc})$ with a normalized Gaussian. 

\section{Moments of the intensity radiation field in the observer reference frame}
\label{appendix_b}

Let us consider a fluid moving with bulk velocity $\vec{V}$ with respect to a static system
of reference, i.e. the observer reference frame.
In the diffusion approximation, the specific intensity in the fluid reference frame can be written
as \citep[e.g.][]{pomraning73}
\begin{equation}
\label{edd_a}
I_{\rm f}= \frac{I_{\rm 0,f}}{4 \pi}   + \frac{3}{4 \pi} \vec{k_f} \cdot \vec{M}_{\rm f},
\end{equation}
where $\vec{k_f}$ is the cosine-direction versor, while  $I_{\rm 0,f}$ and $\vec{M}_{\rm f}$ are the zeroth and first 
moments of the specific intensity, respectively. The latter are defined as
\begin{equation}
\label{zero_moment}
I_{\rm 0,f}= \int I_{\rm f} d\Omega_{\rm f},
\end{equation}
and
\begin{equation}
\label{first_moment}
M^{\rm i}_{\rm f}= \int I_{\rm f} k^{\rm i}_{\rm f} d\Omega_{\rm f}, 
\end{equation}
respectively. 
Note that $M^{\rm i}_{\rm f}$ is the spectral flux along the $i^{\rm th}_{\rm f}$-coordinate. 
Together with the zeroth and first moment, one can also define the second moment of the radiation field
\begin{equation}
\label{second_moment}
K^{\rm ij}_{\rm f}= \int I_{\rm f} k^{\rm i}_{\rm f} k^{\rm j}_{\rm f} d\Omega_{\rm f}.
\end{equation}

For the sake of clarity, we consider here the case of a cartesian coordinate system
where only the $\mz$ component is different from zero (slab approximation) so that equation 
(\ref{edd_a}) can be written as
\begin{equation}
\label{edd_zcomp}
I_{\rm f}=\frac{I_{\rm 0,f}}{4 \pi}   + \frac{3}{4 \pi} \uf \mz,
\end{equation}
where $k_{\rm f}^{\rm z}=u_{\rm f}$ is the cosine of the photon direction along the $z$ coordinate.
Substituting $I_{\rm f}$ from equation (\ref{edd_zcomp}) into the integrals (\ref{zero_moment}) to (\ref{second_moment}) 
it is straightforward to see that 
$K_{\rm f}/I_{\rm 0,f}=1/3$, where $K_{\rm f}=K^{\rm zz}_{\rm f}$.

We now consider the relation between the zeroth and second moment in
the observer reference frame.
The transformation law of the photon energy between the fluid and observer reference frame is
\begin{equation}
\label{doppler}
\el=\ef \Gamma (1+ \beta \uf),
\end{equation} 
where $\beta=V/c$ is the fluid velocity, and the relation between the angles
$\uf$ and $\ul$ is given by the aberration formula
\begin{equation}
\label{aberration}
\uf=\frac{\ul - \beta}{1- \beta \ul}.
\end{equation}
Using now the relativistic invariant 
\begin{equation}
\frac{I_{\rm l}}{\el^3}=\frac{I_{\rm f}}{\ef^3},
\end{equation}
together with the relation (\ref{doppler}), the specific intensity in the laboratory 
reference frame can be written as
\begin{equation}
\label{edd_labframe}
I_{\rm l}=\left[\Gamma (1+ \beta \uf)\right]^3 I_{\rm f},
\end{equation}
where $I_{\rm f}$ is given in equation (\ref{edd_a}).
Computing the zeroth, first, and second moment of the specific intensity in the observer reference 
frame we obtain
\begin{equation}
\label{zero_lab}
J_{\rm l}=\Gamma (I_{\rm 0,f}+ \mz_{\rm f} \beta), 
\end{equation}
\begin{equation}
\label{first_lab}
M^{\rm z}_{\rm l}=\Gamma (\mz_{\rm f}+ I_{\rm 0,f} \beta), 
\end{equation}
and
\begin{eqnarray}
\label{second_lab}
K^{\rm zz}_{\rm l}=\frac{\Gamma}{\beta^3 \Gamma^3}  \left[I_{\rm 0,f} \beta (-1 + 2 \beta^2) + 
    \mz_{\rm f} (3 - 5 \beta^2 + 3 \beta^4)\right]   \\\nonumber 
    +  \frac{(I_{\rm 0,f} \beta -3 \mz_{\rm f} ) {\rm ArcTanh}[\beta]}{\beta^4 \Gamma^3}.
\end{eqnarray}
We also made use above of the equations (\ref{edd_labframe}) and (\ref{aberration}). 
From equations (\ref{zero_lab}) and (\ref{second_lab}) we finally obtain
\begin{equation}
\frac{K^{\rm zz}_{\rm l}}{J_{\rm l}} \approx \frac{1}{3}+\frac{16 \mz_{\rm f} \beta}{15 I_{\rm 0,f}}+ {\rm O}\left[\beta^2\right]
\end{equation}
\end{appendix}

\section*{Acknowledgements}
The authors thank the anonymous referee who provided many suggestions to significantly
improve the first version of the paper.
They are also grateful to the ASI data centre for preserving and making available the \sax 
data, and to F. F\"urst  for providing the \nustar\ and \suzaku\ spectra of \her used in this work.
This research has made use of data obtained with the NuSTAR mission, a project led by the California 
Institute of Technology (Caltech), managed by the Jet Propulsion Laboratory (JPL) and funded by NASA, 
of observations obtained with the \suzaku satellite, a collaborative mission between the space agencies 
of Japan (JAXA) and the USA (NASA), and with the \mbox{X-ray} astronomy satellite BeppoSAX, a project of the 
Italian Space Agency (ASI) with participation of the Netherlands Agency for Aerospace Programs (NIVR).

\bibliographystyle{aa}
\bibliography{compmag2}

\end{document}